\documentclass[fleqn,usenatbib,useAMS]{Papers}
\usepackage{newtxtext,newtxmath}
\usepackage[T1]{fontenc}
\usepackage{subfloat}
\usepackage{makecell}
\usepackage[authoryear]{natbib}
\usepackage{graphicx}	
\usepackage{amsmath}	

\title[BHMR from mass and energy cascade in dark matter flow]{The baryonic-to-halo mass relation from mass and energy cascade in self-gravitating collisionless dark matter flow}

\author[Z. Xu]{
Zhijie (Jay) Xu,$^{1}$\thanks{E-mail: zhijie.xu@pnnl.gov; zhijiexu@hotmail.com}
\\
$^{1}$Physical and Computational Sciences Directorate, Pacific Northwest National Laboratory; Richland, WA 99352, USA\\
}

\date{Accepted XXX. Received YYY; in original form ZZZ}

\pubyear{2022}

\begin{document}
\label{firstpage}
\pagerange{\pageref{firstpage}--\pageref{lastpage}}
\maketitle

\begin{abstract}
The relation between properties of galaxies and dark matter (DM) halos they reside in can be valuable to understand the structure formation and evolution. In particular, the baryonic-to-halo mass ratio (BHMR) and its evolution may provide many important insights. We first review unique properties of self-gravitating collisionless dark matter flow (SG-CFD), followed by their applications in deriving BHMR. To maximize system entropy, the long-range interaction requires a broad size of halos to be formed. These halos facilitate an inverse mass and energy cascade from small to large scales that involves a constant rate of energy cascade $\varepsilon_u \approx -4.6\times 10^{-7} m^2/s^3$. The mass and energy cascade represent an intermediate statistically steady state of dark matter flow. In addition, dark matter flow exhibits scale-dependent flow behaviors that is incompressible on small scale and irrotational on large scale. Considering a given halo with a total baryonic mass $m_b$, halo mass $m_h$, halo virial size $r_h$, and flat rotation speed $v_f$, the baryonic-to-halo mass relation can be analytically derived by combining the baryonic Tully-Fisher relation and the rate of energy cascade $\varepsilon_u$ in small and large halos. We found a maximum BHMR ratio ~0.076 for halos with a critical mass $m_{hc}\sim 10^{12}M_{\odot}$ at z=0. That ratio is much lower for both smaller and larger halos such that two regimes can be identified: i) for incompressible small halos with mass $m_h<m_{hc}$, we have $\varepsilon_u \propto v_f/r_h$, $v_f \propto r_h$, and $m_b \propto \left(m_{h} \right)^{{4/3} } $; ii) for large halos with mass $m_h>m_{hc}$, we have $\varepsilon_u \propto {v_f^3/r_h}$, $v_f\propto r_h^{1/3}$, and $m_b\propto(m_h)^{4/9}$. Combined with double-$\lambdaup$ halo mass function, the average BHMR ratio in all halos (~0.024 at z=0) can be analytically derived, along with its redshift evolution. The fraction of total baryons in all galaxies is ~7.6\% at z=0 and increases with time $\propto t^{1/3}$. The SPARC (Spitzer Photometry \& Accurate Rotation Curves) data with 175 late-type galaxies were used for derivation and comparison. 
\end{abstract}

\begin{keywords}
\vspace*{-10pt}
Dark matter; N-body simulations; Baryonic mass; Stellar-to-halo ratio;
\end{keywords}

\begingroup
\let\clearpage\relax
\tableofcontents
\endgroup
\vspace*{-20pt}

\section{Introduction}
\label{sec:1}
The dark matter problem originates from the mass discrepancy between required amount from Newtonian gravity (dynamical mass) and the directly observed amount of baryonic mass. The flat rotation curves of spiral galaxies directly point to this discrepancy: total mass predicted by Newtonian gravity is much greater than the observed mass from luminous matter \citep{Rubin:1970-Rotation-of-Andromeda-Nebula-f,Rubin:1980-Rotational-Properties-of-21-Sc}. The standard $\Lambda$CDM model interprets this mismatch by postulating the existence of dark matter halos that galaxies reside in. Thus, the baryonic-to-halo mass relation (BHMR) represents a fundamental relation to describe this mass discrepancy. The total baryonic mass is defined as the sum of the mass from stars (stellar mass) and cold gas. The stellar mass is intimately coupled to the depth of halo potential, and thus to the mass of halos. The stellar-to-halo mass relation (SHMR) should be related to the BHMR and reflect the accretion and feedback processes in galaxy formation \citep{Wechsler:2018-The-Connection-Between-Galaxie}.

The stellar-to-halo mass ratio is the greatest in halos with a critical size $m_{hc}\sim10^{12}M_{\odot}$ and is much smaller for both smaller and larger halos \citep{Moster:2013-Galactic-star-formation-and-ac,Moster:2010-Constraints-on-the-Relationshi,Girelli:2020-The-stellar-to-halo-mass-relat}. The empirical SHMR can be obtained by a halo abundance matching approach, in which galaxy properties can be linked to dark matter halos from N-body simulations \citep{Guo:2010-How-do-galaxies-populate-dark-,Behroozi:2010-A-COMPREHENSIVE-ANALYSIS-OF-UN}. Dark matter (DM), if exists, is believed to be cold, collisionless, dissipationless, non-baryonic, and barely interacting with baryonic matter except through gravity. In addition, dark matter should be sufficiently smooth on large scales with a fluid-like behavior, i.e. the dark matter flow that is best described by self-gravitating collisionless fluid dynamics (SG-CFD). This paper focus on the derivation of BHMR that can be determined from the nature of dark matter flow.

First, the baryonic mass $m_{b}$ can be related to the flat rotation velocity $v_{f} $ through a baryonic Tully-Fisher relation $v_{f}^{4} =Gm_{b} a_{0} $ (BTFR), where $a_{0} \approx 1.2\times 10^{-10} {m/s^{2} } $ is an empirical constant of acceleration and \textit{G} is the gravitational constant. This empirical relation was originally established for rotation velocity \citep{Tully:1977-New-Method-of-Determining-Dist}. Similar scaling was also found for the velocity dispersion $\sigma $ of stars \citep{Faber:1976-Velocity-Dispersions-and-Mass-}, i.e. $\sigma \propto \left(m_{b} \right)^{{1/4} } $. The BTFR is a natural result of the modified Newtonian Dynamics (MOND) \citep{Milgrom:1983-A-Modification-of-the-Newtonia}. As an ad hoc empirical theory, MOND successfully explains the shape of rotation curves \citep{McGaugh:1998-Testing-the-dark-matter-hypoth}, the baryonic Tully-Fisher relation \citep{McGaugh:2000-The-baryonic-Tully-Fisher-rela,Lelli:2019-The-baryonic-Tully-Fisher-rela}, and many others. This motivates the search of a fundamental theory explaining the MOND paradigm. 

One potential theory is to consider MOND as an effective theory describing the dynamics of baryonic mass suspended in dark matter flow that involves fluctuations in both velocity and acceleration with typical scales $u$ and $a_0$ \citep{Xu:2022-The-origin-of-MOND-acceleratio}. In dark matter flow, the halo-mediated inverse energy cascade from small to large mass sales involves a constant rate of energy cascade $\varepsilon _{u} \approx -4.6\times 10^{-7} {m^{2}/s^{3}}$ that can be related to both fluctuations as $\varepsilon _{u} =-{a_{0}u/(3\pi) ^{2}}$, where factor $3\pi$ is from the angle of incidence. With $u_{0} \equiv u(z=0)\approx 354.61{km/s}$ from N-body simulation, the scale of acceleration fluctuation $a_{0} \left(z=0\right)\approx 1.2\times 10^{-10} {m/s^{2}}$ can be easily obtained. The acceleration fluctuation seems successfully explain the origin of critical MOND acceleration $a_{0}$. In this regard, the BTFR might also be a direct manifestation of the fluctuating dark matter flow.

Second, the halo mass $m_{h} $ can be related to the halo virial radius $r_{h} $ with a scaling $m_{h} \propto \left(r_{h} \right)^{3} $ (Eq. \eqref{ZEqnNum246288}). This can be obtained from a spherical collapse model \citep{Gunn:1972-Infall-of-Matter-into-Clusters} or a two-body collapse model or TBCM \citep{Xu:2021-A-non-radial-two-body-collapse}, where a constant density ratio $\Delta _{c}=18\pi^2$ between mean halo density and background density can be identified such that $m_{h} \propto \Delta _{c} \left(r_{h} \right)^{3}$. 

Finally, the BHMR (the relation between $m_{b} $ and $m_{h} $) can be obtained only if the relation between flat rotation velocity $v_{f} $ and halo virial radius $r_{h} $ is known. This relation may be determined from the mass and energy cascade in dark matter flow.

\begin{figure}
\includegraphics*[width=\columnwidth]{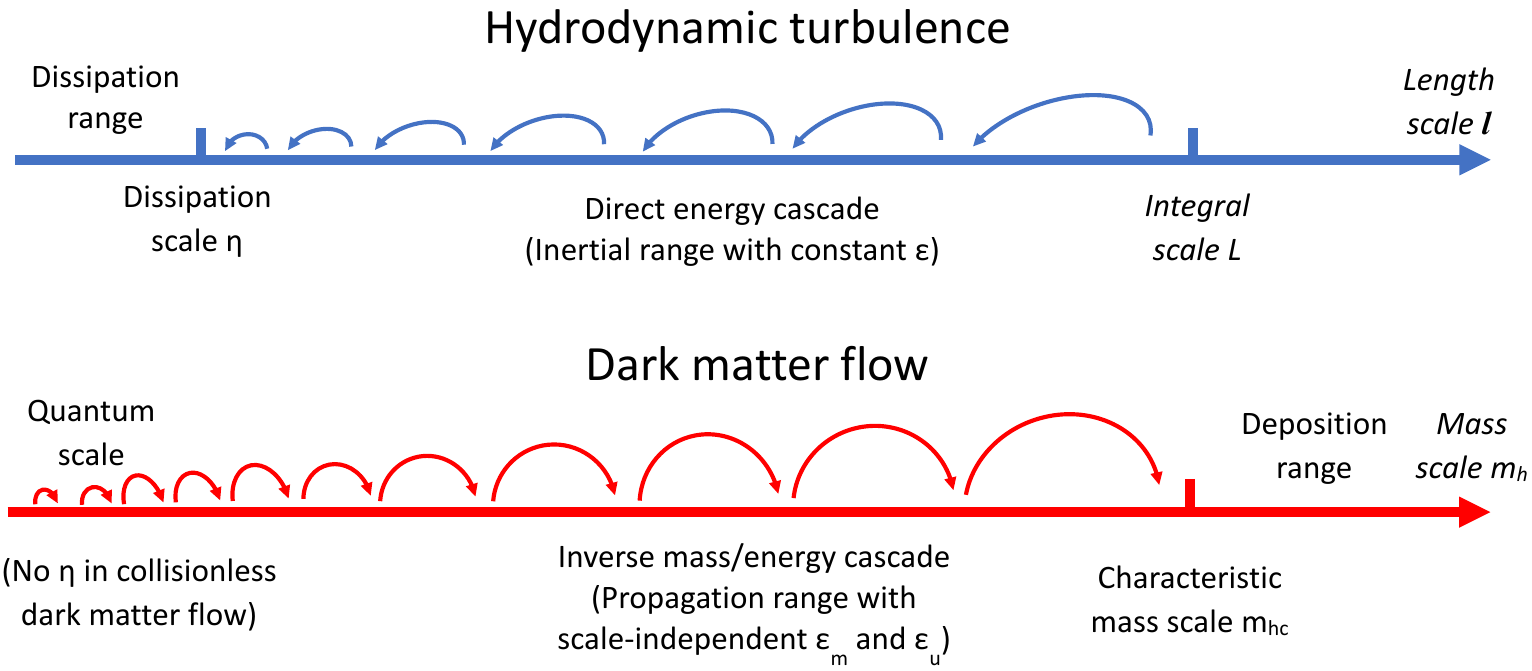}
\caption{Schematic plot of direct energy cascade in hydrodynamic turbulence that is mediated by eddies of different scales. By contrast, halos of different sizes facilitate the inverse mass/energy cascade in dark matter flow.} 
\label{fig:S1}
\end{figure}

Apparently, both dark matter flow and hydrodynamic turbulence share some common features including the randomness, nonlinearity, and multiscale nature \citep{Xu:2022-Dark_matter-flow-and-hydrodynamic-turbulence-presentation}. The homogeneous isotropic incompressible turbulence has been well-studied for many decades \citep{Taylor:1935-Statistical-theory-of-turbulan,Taylor:1938-Production-and-dissipation-of-,de_Karman:1938-On-the-statistical-theory-of-i,Batchelor:1953-The-Theory-of-Homogeneous-Turb}. Turbulence consists of eddies (the building blocks) with a typical velocity \textit{u} and size \textit{l} on different scales. The classical picture of turbulence is an eddy-mediated cascade process (Fig. \ref{fig:S1}), where the kinetic energy of large eddies feeds smaller eddies, which feeds even smaller eddies, and so on to the smallest scale $\eta $ (dissipation scale). Below scale $\eta $, the viscous dissipation becomes dominant (i.e. the dissipation range). Greater than scale $\eta$, there exists a range of length scales with a scale-independent constant rate $\varepsilon $ (unit: ${m^{2} /s^{3} } $) at which energy is passing down the cascade (i.e. the inertial range): 
\begin{equation} 
\label{ZEqnNum629737} 
\varepsilon \approx A_{0} \frac{u^{2} }{\left({l/u} \right)} ,            
\end{equation} 
where the kinetic energy is transferred by the amount of $u^{2} $ in a period of eddy turnaround time ${l/u} $. Here $A_{0} $ is a numerical constant. The smallest (Kolmogorov) length scale $\eta = ({\nu ^{3} /\varepsilon })^{{1/4} } $ is determined by both $\varepsilon $ and fluid viscosity $\nu $. This process, a direct (kinetic) energy cascade, can be best described by a famous poem \citep{Richardson:1922-Weather-Prediction-by-Numerica}: 

\bigbreak
\centerline{"Big whirls have little whirls, That feed on their velocity;}
\centerline{And little whirls have lesser whirls, And so on to viscosity."} 
\bigbreak

However, the dark matter flow is unique because of its collisionless nature and long-range interaction \citep{Xu:2022-Dark_matter-flow-and-hydrodynamic-turbulence-presentation}. First, the long-range interaction leads to the fluctuation in acceleration that might explain the critical MOND acceleration $a_0$ \citep{Xu:2022-The-origin-of-MOND-acceleratio}. In addition, a broad size of halos are required to be formed to maximize system entropy in systems with long-range interactions \citep{Xu:2021-The-maximum-entropy-distributi,Xu:2021-Mass-functions-of-dark-matter-}. These halos (counterpart to "eddies" in turbulence) facilitate an inverse mass cascade from small to large mass scales that is not present in hydrodynamic turbulence (Fig. \ref{fig:S1}). The inverse mass cascade leads to new understandings of halo mass functions and internal structures (density profiles etc.) based on the random-walk of halos in mass space and random-walk of particles in size-varying halos \citep{Xu:2021-Inverse-mass-cascade-halo-density,Xu:2021-Inverse-mass-cascade-mass-function}. From this description, the mass cascade in dark matter flow can be described by a similar poem with "whirls" replaced by "halos": 

\bigbreak
\centerline{"Little halos have big halos, That feed on their mass;} 
\centerline{And big halos have greater halos, And so on to growth."} 
\bigbreak

Note that both turbulence and dark matter flow are non-equilibrium systems involve energy cascade across different scales (Fig. \ref{fig:S1}) \citep{Xu:2021-Inverse-and-direct-cascade-of-}. The mass/energy cascade is an essential feature of intermediate statistically steady state for non-equilibrium systems to continuously maximize system entropy. Inspired by Eq. \eqref{ZEqnNum629737}, the constant rate of energy transfer $\varepsilon _{u} $ in dark matter flow might be similarly determined by the flat rotation speed and halo virial size as  
\begin{equation} 
\label{ZEqnNum644698} 
\varepsilon _{u} \propto \frac{v_{f}^{2} }{\left({r_{h} /v_{f} } \right)} ,           
\end{equation} 
which provides a relation between $v_{f}$ and $r_{h}$ and eventually facilitates the derivation of BHMR. 

However, unlike hydrodynamic turbulence that is incompressible on all scales, dark matter flow is much more complex. It exhibits scale-dependent flow behaviors, i.e. a constant divergence flow for peculiar velocity on small scales and an irrotational flow on large scales \citep{Xu:2022-The-statistical-theory-of-2nd,Xu:2022-Two-thirds-law-for-pairwise-ve,Xu:2022-The-statistical-theory-of-3rd}. This hints that Eq. \eqref{ZEqnNum644698} might be good only for a certain range of halo size. In fact, it is good for large halos with mass $m_{h}>m_{hc}$ (Eq. \eqref{ZEqnNum886640}), where $m_{hc}$ is the critical halo size with the greatest baryonic-to-halo mass ratio (see Eq. \eqref{ZEqnNum832262}).

In addition, the collisionless nature means the absence of viscous dissipation in dark matter flow such that the smallest length scale for inertial range ($\eta$) is not limited by viscosity (Fig. \ref{fig:S1}). This enables us to extend the scale-independent constant $\varepsilon_u$ down to the smallest scale where quantum effects are important, such that dark matter particle mass, size and other properties might be predicted \citep{Xu:2022-Postulating-dark-matter-partic}. 

In this paper, we first review some unique properties of dark matter flow, followed by the derivation of baryonic-to-halo mass relation based on these properties. 

\section{N-body simulations for dark matter flow}
\label{sec:2}
The basic dynamics of dark mater flow is governed by the collisionless Boltzmann equations (CBE) \citep{Mo:2010-Galaxy-formation-and-evolution} that can be numerically solved by particle-based N-body simulations \citep{Peebles:1980-The-Large-Scale-Structure-of-t}. The simulation data for this work was generated from large scale N-body simulations by the Virgo consortium \citep{Frenk:2000-Public-Release-of-N-body-simul,Jenkins:1998-Evolution-of-structure-in-cold}. The current work focuses on the matter-dominant simulations with $\Omega _{0} =1$ and cosmological constant $\Lambda =0$. The same set of simulation data has been widely used in various studies such as the clustering statistics \citep{Jenkins:1998-Evolution-of-structure-in-cold}, the formation of halo clusters in large scale environments \citep{Colberg:1999-Linking-cluster-formation-to-l}, and testing models for halo abundance and mass functions \citep{Sheth:2001-Ellipsoidal-collapse-and-an-im}. Key parameters of N-body simulations are listed in Table \ref{tab:1}, where \textit{h} is the Hubble constant in the unit of $100{km/\left(Mpc\cdot s\right)} $, $N$ is the number of particles, and $m_{p} $ is the particle mass. 

\begin{table}
\caption{Numerical parameters of N-body simulation}
\begin{tabular}{p{0.25in}p{0.05in}p{0.05in}p{0.05in}p{0.05in}p{0.05in}p{0.4in}p{0.1in}p{0.35in}p{0.35in}} 
\hline 
Run & $\Omega_{0}$ & $\Lambda$ & $h$ & $\Gamma$ & $\sigma _{8}$ & \makecell{L\\(Mpc/h)} & $N$ & \makecell{$m_{p}$\\$M_{\odot}/h$} & \makecell{$l_{soft}$\\(Kpc/h)} \\ 
\hline 
SCDM1 & 1.0 & 0.0 & 0.5 & 0.5 & 0.51 & \centering 239.5 & $256^{3}$ & 2.27$\times 10^{11}$ & \makecell{\centering 36} \\ 
\hline 
\end{tabular}
\label{tab:1}
\end{table}

Friends-of-friends algorithm (FOF) was used to identify all halos in simulation that depends only on a dimensionless parameter \textit{b}, which defines the linking length $b\left({N/V} \right)^{{-1/3} } $, where $V$ is the volume of simulation box. All halos in simulation were identified with a linking length parameter of $b=0.2$ in this work. All halos identified were grouped into halo groups of different size according to halo mass $m_{h} $ (or $n_{p} $, the number of particles in halo), where $m_{h} =n_{p} m_{p} $. The total mass for a group of halos of mass $m_{h} $ is the group mass $m_{g} =m_{h} n_{h} $, where $n_{h} $ is the number of halos in that group. Simulation results are presented to describe the mass/energy cascade across halo groups of different mass scales. The relevant dark flow datasets using this simulation are also provided \citep{Xu:2022-Dark_matter-flow-dataset-part1,Xu:2022-Dark_matter-flow-dataset-part2,Xu:Dark_matter_flow_dataset_2022_all_files}. All data files are also available on GitHub \citep{Xu:Dark_matter_flow_dataset_2022_all_files}.

\section{Mass cascade in dark matter flow}
\label{sec:3}
\begin{figure}
\includegraphics*[width=\columnwidth]{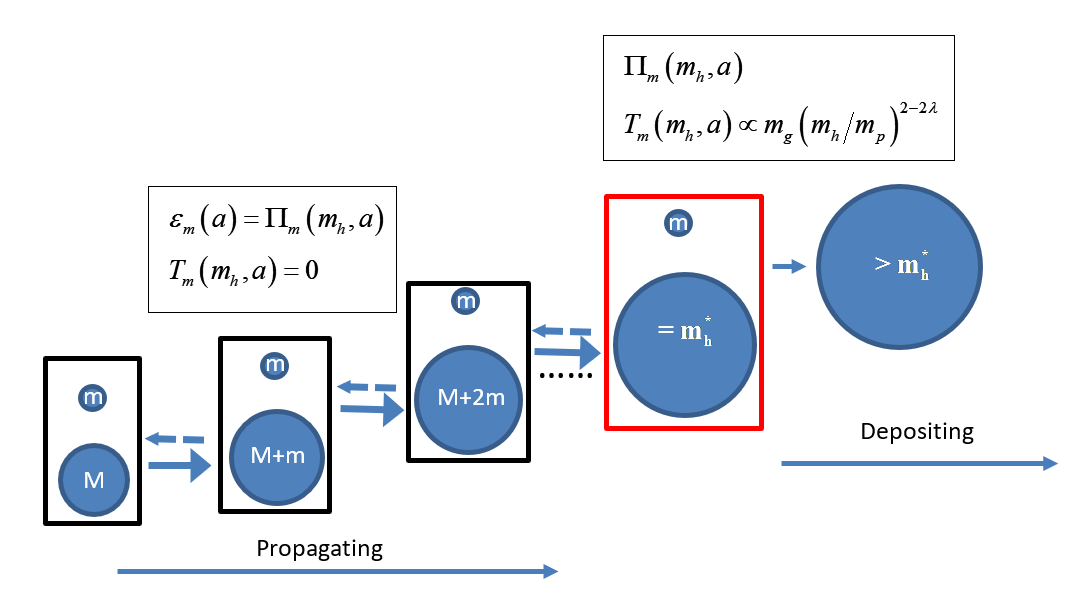}
\caption{Schematic plot of a chain reaction description for inverse mass cascade. Halos merge with free radicals (single mergers) to cause the next merging along the chain and facilitate a continuous mass cascade along the reaction chain. Mass flux function $\Pi _{m} \left(m_h,a\right)$ can be computed using Eq. \eqref{ZEqnNum986318}. A scale-independent mass flux $\varepsilon _{m} \left(a\right)$ is expected for halos smaller than the characteristic mass ($m_{h} <m_{h}^{*} $, i.e. propagation range). Mass cascaded from small scales is consumed to grow halos at scales $m_{h} >m_{h}^{*} $ with a scale-dependent mass flux $\Pi _{m} \left(m_{h} ,a\right)$ (deposition range).} 
\label{fig:1}
\end{figure}

Dark matter flow exhibits unique behaviors due to its collisionless and long-range interaction nature. The highly localized and over-dense halos are results of nonlinear gravitational collapse \citep{Neyman:1952-A-Theory-of-the-Spatial-Distri,Cooray:2002-Halo-models-of-large-scale-str} and the building blocks of SG-CFD. Halos facilitate an inverse mass cascade (Fig. \ref{fig:S1}) that is not present in regular turbulence \citep{Xu:2021-Inverse-mass-cascade-mass-function}. Figure \ref{fig:1} provides a simple chain reaction description for inverse mass cascade that is local in mass space, two-way (forward/backward, solid/dash arrows in Fig. \ref{fig:1}), and asymmetric process (forward is dominant over backward). The net mass transfer proceeds in a "bottom-up" fashion from small to large mass scales (inverse cascade). Halos pass their mass onto larger and larger halos, until halo mass growth becomes dominant over mass propagation for halos with a mass $m_h>m_h^*$. 

The mass cascade generally involves three stages \citep{Xu:2021-Inverse-mass-cascade-mass-function}: 
\begin{enumerate}
\item \noindent The initial stage corresponds to the generation of the single mergers (free radicals) to provide the mass and energy source for halos; 
\item \noindent The propagation stage for halos with a mass $m_{h} <m_{h}^{*} $ involving a sequence of merging between single mergers and halos to propagate mass along the reaction chain (a propagation range, counterpart to the "inertial range" in turbulence). The rate of mass transfer from scale below $m_{h} $ to above $m_{h} $ is independent of $m_{h}$ in this range, 
\begin{equation}
\varepsilon_{m} \left(a\right)=\Pi _{m} \left(m_{h},a\right)=-m_{h} f_{h} \left(m_{h},a\right) \quad \textrm{for} \quad m_{h} \ll m_{h}^{*},  
\label{ZEqnNum379703}
\end{equation}
where the merging frequency $f_{h} \left(m_{h} ,a\right)$ for a group of halos of the same mass $m_{h}$ is proportional to the number of halos $n_{h} $ (term 1 in Eq. \eqref{ZEqnNum596894}) in that halo group and the halo surface area (term 2 with a geometry parameter $\lambda \approx {2/3} $ for surface area),  
\begin{equation} 
\label{ZEqnNum596894} 
f_{h} \left(m_{h} ,a\right)=f_{0} \left(a\right)\underbrace{M_{h} \left(a\right)f_{M} \left(m_{h} ,m_{h}^{*} \right)\frac{m_{p} }{m_{h} } }_{1}\underbrace{\left(\frac{m_{h} }{m_{p} } \right)^{\lambda } }_{2},     
\end{equation} 
where $f_{0} \left(a\right)\propto a^{-1} $ is a fundamental frequency for the merging between two single mergers \citep{Xu:2021-Inverse-mass-cascade-mass-function}. 

The halo mass function $f_{M} \left(m_{h} ,m_{h}^{*} \left(a\right)\right)$ is the probability distribution of total halo mass $M_{h} \left(a\right)$ with respect to the halo mass $m_{h} $. The more halos in a group and the larger halo surface area will have a greater probability (greater $f_{h} $) for halos to merge with a single merger. After substituting Eq. \eqref{ZEqnNum596894}, a dimensional analysis for a scale-independent $\varepsilon _{m} \left(a\right)$ in Eq. \eqref{ZEqnNum379703} leads to a simple power-law halo mass function in propagation range, 
\begin{equation}
f_{M} \left(m_{h} ,m_{h}^{*} \right)\propto m_{h}^{-\lambda } \left(m_{h}^{*} \right)^{\lambda -1} \quad \textrm{for} \quad m_{h} <m_{h}^{*}.  
\label{ZEqnNum972525}
\end{equation}
\item \noindent The termination stage ($m_{h} >m_{h}^{*} $) involves the deposition of the mass cascaded from the scales below $m_{h}^{*} $ to scales larger than $m_{h}^{*} $ (a deposition range, counterpart to the ``dissipation range'' in turbulence). The dominant mode at this stage is the growth of halos with a scale-dependent mass flux function $\Pi _{m} \left(m_{h} ,a\right)$. 
\end{enumerate}

The real-space mass flux function can be introduced to quantify the net transfer of mass from all halos smaller than scale $m_{h} $ to all halos greater than $m_{h} $,
\begin{equation} 
\label{ZEqnNum986318} 
\Pi _{m} \left(m_{h} ,a\right)=-\frac{\partial }{\partial t} \left[M_{h} \left(a\right)\int _{m_{h} }^{\infty }f_{M} \left(m,m_{h}^{*} \right) dm\right].       
\end{equation} 
The mass transfer function can be introduced to quantify the rate of change of group mass $m_{g} \left(m_{h} ,a\right)=n_{h} m_{h} $,
\begin{equation}
\label{ZEqnNum400994} 
T_{m} \left(m_{h} ,a\right)=\frac{\partial \Pi _{m} \left(m_{h} ,a\right)}{\partial m_{h} } =\frac{\partial m_{g} \left(m_{h} ,a\right)}{m_{p} \partial t} .       
\end{equation} 
Here $M_{h} \left(a\right)\propto a^{{1/2} }$ is the total mass in all halos and $M_{h} \left(z=0\right)\approx 0.6Nm_{p} $ from N-body simulation \citep{Xu:2021-Inverse-mass-cascade-mass-function,Xu:2021-Inverse-and-direct-cascade-of-}. The total DM mass in halos is about 60\% of all DM and the rest 40\% DM is in single mergers that do not belong to any halos (out-of-halo). This information can be used to estimate the fraction of total baryonic mass in halos and out-of-halo (see Eq. \eqref{ZEqnNum269995}). 

\begin{figure}
\includegraphics*[width=\columnwidth]{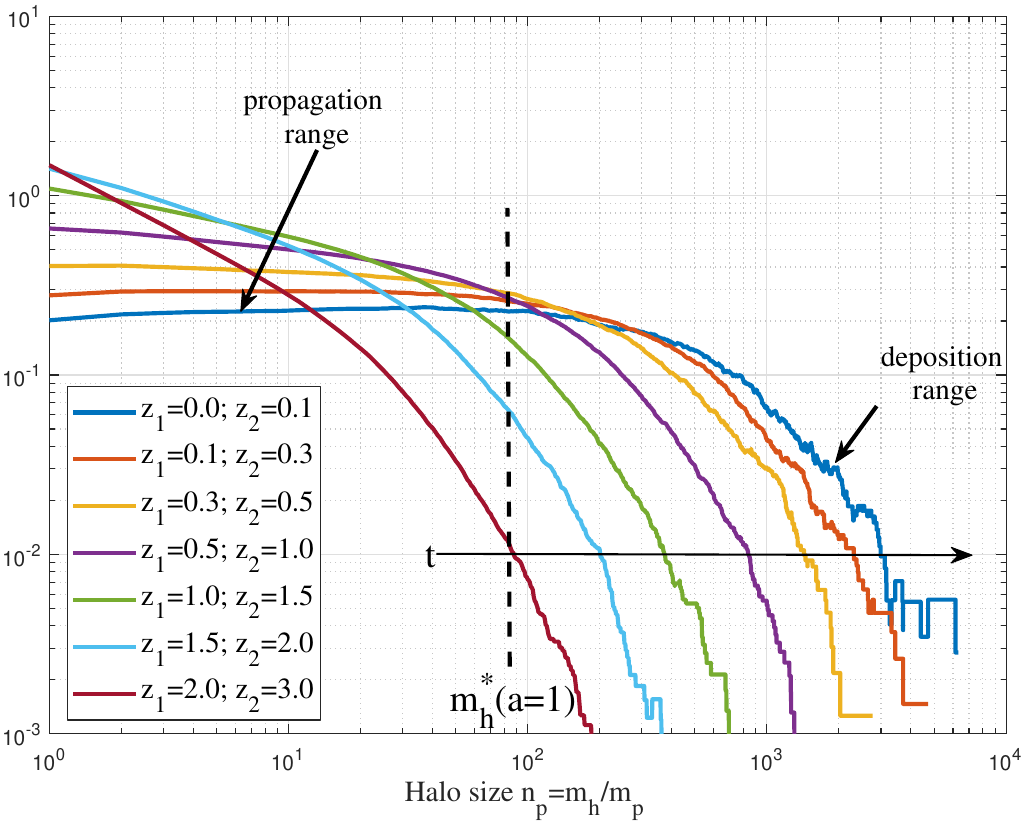}
\caption{The mass flux function $-\Pi _{m} \left(m_{h} ,a\right)$ (normalized by ${Nm_{p} /t_{0} } $) computed from \textit{N}-body simulation at two different redshifts \textit{z$_{1}$ }and \textit{z$_{2}$} (using Eq. \eqref{ZEqnNum986318} and dark matter flow dataset \citep{Xu:2022-Dark_matter-flow-dataset-part1}), where $m_{h} =n_{p} m_{p} $. A scale-independent mass flux $\varepsilon _{m} \left(a\right)$ can be found for halo groups smaller than the characteristic mass $m_{h} <m_{h}^{*} $. The negative mass flux indicates an inverse mass cascade from small to large mass scales. The propagation and deposition ranges can be clearly identified.}
\label{fig:2}
\end{figure}

Figure \ref{fig:2} presents the variation of mass flux functions with halo size computed by Eq. \eqref{ZEqnNum986318} and dark matter flow dataset \citep{Xu:2022-Dark_matter-flow-dataset-part1}. Two distinct ranges can be clearly identified, i.e. a propagation range with scale-independent $\varepsilon _{m} \sim a^{-1} $ for $m_{h} <m_{h}^{*} $ and a deposition range with cascaded mass consumed to grow halos for $m_{h} >m_{h}^{*} $. The scale-independent mass flux $\varepsilon _{m} $ can be finally written as \citep[see][Eq. (54)]{Xu:2021-Inverse-mass-cascade-mass-function}),
\begin{equation} 
\label{eq:8} 
\varepsilon _{m} \left(a\right)=-{M_{h} \left(a\right)H/2} =-{M_{h} \left(a\right)/\left(3t\right)} \propto a^{-1} .      
\end{equation} 

Based on the chain reaction description in Fig. \ref{fig:1}, the inverse mass cascade can be further refined in terms of a random walk of halos in mass space. Halos are migrating in mass space (due to merging with single mergers) with a given distribution of waiting time (or the jumping frequency) \citep{Xu:2021-Inverse-mass-cascade-mass-function}. The Press-Schechter (PS) \citep{Bower:1991-The-Evolution-of-Groups-of-Gal,Press:1974-Formation-of-Galaxies-and-Clus} mass function turns out to be a natural result of halo random walk in mass space with a single geometry parameter $\lambdaup$=2/3 for all halos of different size \citep{Xu:2021-Inverse-mass-cascade-mass-function}. A double-$\lambdaup$ mass function is naturally proposed with different $\lambdaup$ for two distinct (propagation/deposition) ranges, which can be conveniently written as \citep[see][Section 5.2]{Xu:2021-Inverse-mass-cascade-mass-function},
\begin{equation} 
\label{ZEqnNum889111} 
f_{D\lambda } \left(\nu \right)=\frac{\left(2\sqrt{\eta _{0} } \right)^{-q_{0} } }{\Gamma \left({q_{0} /2} \right)} \nu ^{{q_{0} /2} -1} \exp \left(-\frac{\nu }{4\eta _{0} } \right),       
\end{equation} 
where the dimensionless variable $\nu=({m_{h} /m_{h}^{*}})^{{2/3}}$. 

\begin{figure}
\includegraphics*[width=\columnwidth]{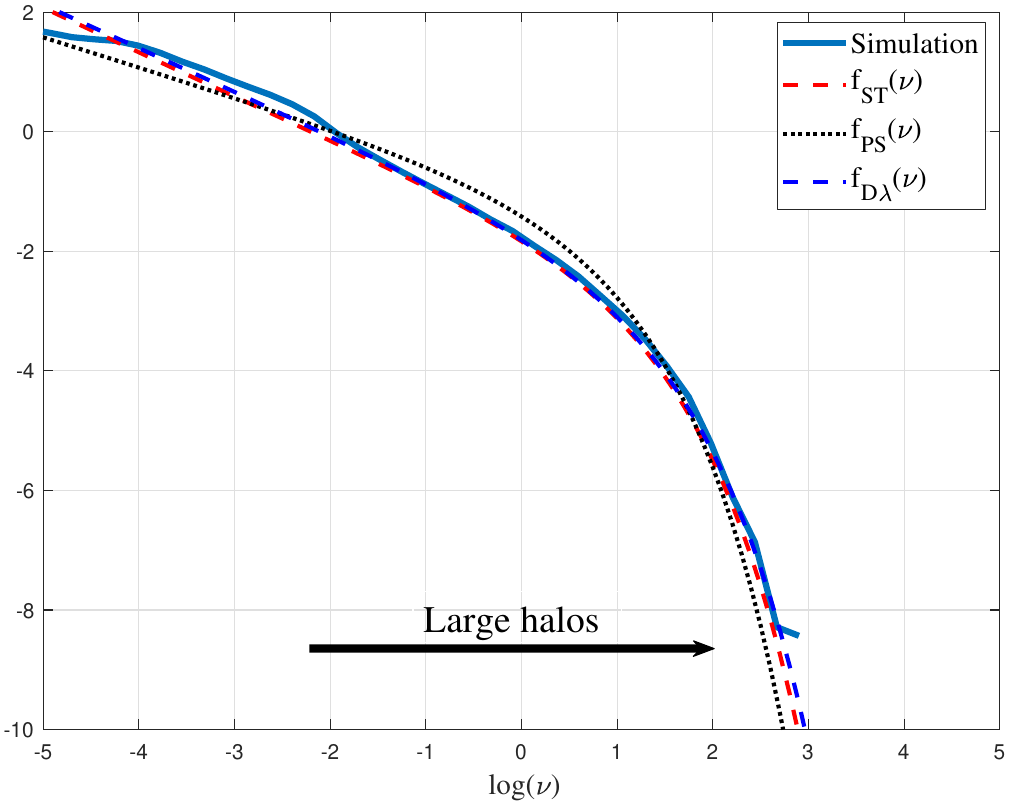}
\caption{Comparison between different mass functions (logf($\nu$)) and simulation data at \textit{z}=0. The PS mass function underestimates the mass in large halos. The double-$\lambdaup$ mass function from Eq. \eqref{ZEqnNum889111} matches both N-body simulation and the ST mass function.}
\label{fig:3}
\end{figure}

Figure \ref{fig:3} presents the comparison between simulation data, double-$\lambdaup$ mass function, PS mass function, and ST mass function from elliptical collapse model \citep{Xu:2021-Inverse-mass-cascade-mass-function,Sheth:2001-Ellipsoidal-collapse-and-an-im}. The values of $\eta _{0} =0.76$ and $q_{0} =0.556$ give the best fit to the simulation data that is much better than the PS mass function. The double-$\lambdaup$ mass function will be used to derive the average baryonic mass fraction in all halos (Eqs. \eqref{ZEqnNum855755} and \eqref{ZEqnNum227267}). 

\section{Energy cascade in dark matter flow}
\label{sec:4}
Though mass cascade is not present in hydrodynamic turbulence, both flows are non-equilibrium systems involving energy cascade across different scales \citep{Xu:2021-Inverse-and-direct-cascade-of-}. To quantify the energy cascade in dark matter flow, peculiar velocity $\boldsymbol{\mathrm{u}}_{p} $ of every halo particle can be decomposed into halo velocity $\boldsymbol{\mathrm{u}}_{h} $ and velocity in halo $\boldsymbol{\mathrm{u}}_{p}^{'}$ \citep[same as][]{Cooray:2002-Halo-models-of-large-scale-str}, i.e. $\boldsymbol{\mathrm{u}}_{p} =\boldsymbol{\mathrm{u}}_{h} +\boldsymbol{\mathrm{u}}_{p}^{'} $, where $\boldsymbol{\mathrm{u}}_{\boldsymbol{\mathrm{h}}} =\left\langle \boldsymbol{\mathrm{u}}_{\boldsymbol{\mathrm{p}}}^{} \right\rangle _{h} $ is the mean velocity of all particles in the same halo. Here $\left\langle \right\rangle _{h} $ stands for the average for all particles in the same halo. The halo virial dispersion $\sigma _{vh}^{2} $ is the variance of velocity $\boldsymbol{\mathrm{u}}_{p}^{'} $ defined for every halo \citep[see][Section 4.1]{Xu:2021-Inverse-and-direct-cascade-of-},
\begin{equation} 
\label{ZEqnNum627513} 
\sigma _{vh}^{2} =\left\langle \left(\boldsymbol{\mathrm{u}}_{p}^{'x} \right)^{2} \right\rangle _{h} =\left\langle \left(\boldsymbol{\mathrm{u}}_{p}^{'y} \right)^{2} \right\rangle _{h} =\left\langle \left(\boldsymbol{\mathrm{u}}_{p}^{'z} \right)^{2} \right\rangle _{h} .        
\end{equation} 

The halo virial dispersion $\sigma _{vh}^{2}$ represents the temperature of that halo. With all halos grouped according to their size, halo groups are characterized by the size of halos in that group ($n_{p} $ or $m_{h} =n_{p} m_{p} $), mean halo virial dispersion ($\sigma _{v}^{2} $), and halo velocity dispersion ($\sigma _{h}^{2} $),
\begin{equation}
\sigma _{v}^{2} =\left\langle \sigma _{vh}^{2} \right\rangle _{g} \textrm{ and } \sigma _{h}^{2} =\left\langle \left(\boldsymbol{\mathrm{u}}_{h}^{x} \right)^{2} \right\rangle _{g} =\left\langle \left(\boldsymbol{\mathrm{u}}_{h}^{y} \right)^{2} \right\rangle _{g} =\left\langle \left(\boldsymbol{\mathrm{u}}_{h}^{z} \right)^{2} \right\rangle _{g},    
\label{ZEqnNum872828}
\end{equation}

\noindent where $\sigma _{h}^{2} $ represents the mean kinetic energy of halos, i.e. the temperature of a halo group. Halo virial dispersion $\sigma _{v}^{2}$ is the mean halo temperature with $\langle\rangle_{g}$ for the average for all halos in the same group. Therefore, the kinetic energy of halo particles can be decomposed
\begin{equation} 
\label{ZEqnNum425919} 
\sigma ^{2} \left(m_{h} ,a\right)=\sigma _{h}^{2} \left(a\right)+\sigma _{v}^{2} \left(m_{h} ,a\right),        
\end{equation} 
where $\sigma _{h}^{2}$ is relatively independent of halo mass $m_{h}$ and $\sigma _{v}^{2} \propto \left(m_{h} \right)^{{2/3} }$ \citep{Xu:2021-Inverse-and-direct-cascade-of-}. 

\begin{figure}
\includegraphics*[width=\columnwidth]{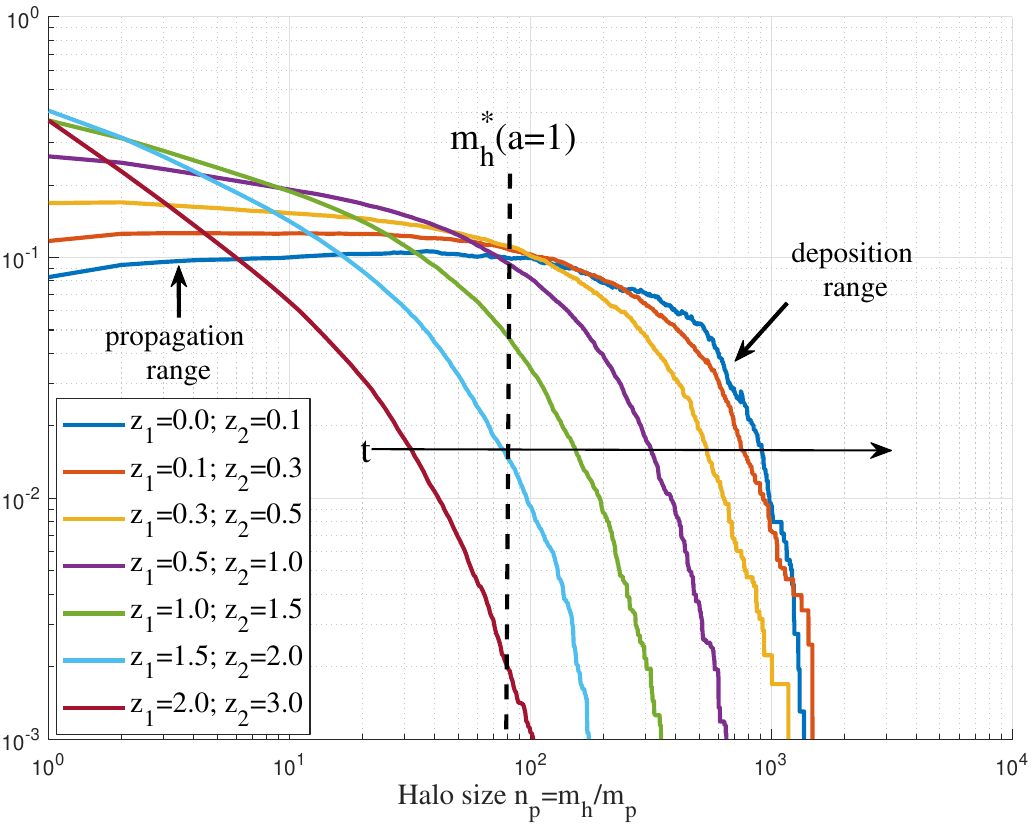}
\caption{The variation of energy flux function $-\Pi _{kh}(m_h,a)$ for halo kinetic energy $\sigma _{h}^{2} $ with the size $n_{p} $ of halo groups. The flux function $\Pi _{kh} <0$ (inverse cascade) is normalized by ${Nm_{p} u_{0}^{2} /t_{0} } $ and computed from simulation results at two different redshifts $z_{1} $ and $z_{2}$ using Eq. \eqref{ZEqnNum299256} and dark matter flow dataset \citep{Xu:2022-Dark_matter-flow-dataset-part1}. A scale-independent flux function $\varepsilon _{kh} $ can be identified for propagation range with $m_{h} <m_{h}^{*} $.}
\label{fig:4}
\end{figure}

\begin{figure}
\includegraphics*[width=\columnwidth]{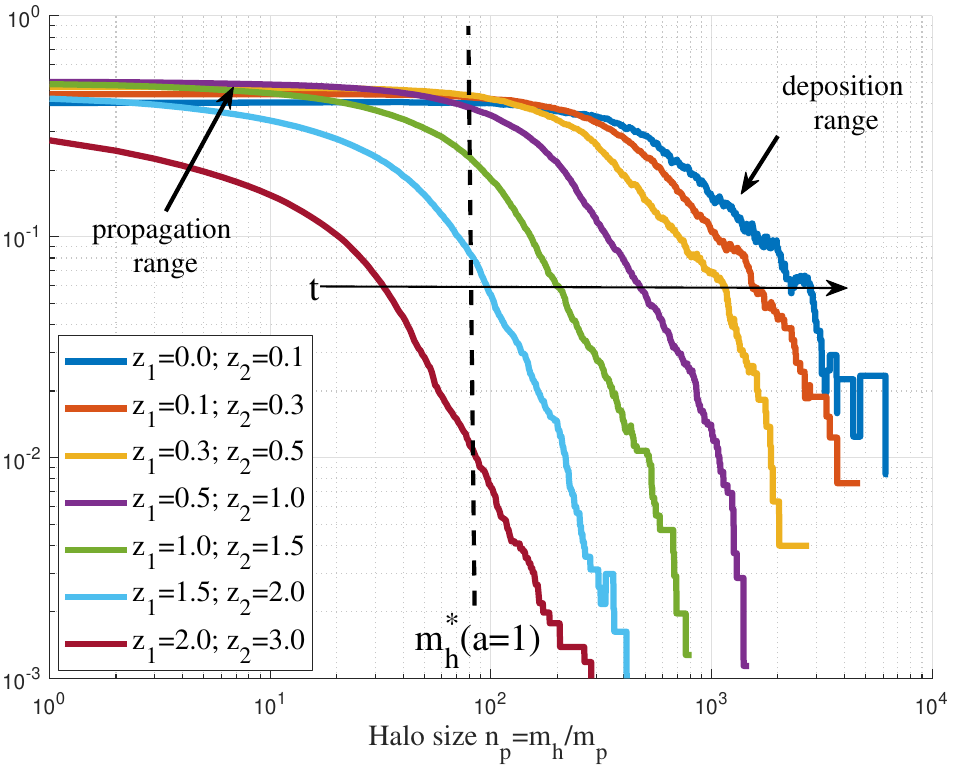}
\caption{The variation of flux function $-\Pi _{kv} \left(m_{h} ,a\right)$ for halo virial energy $\sigma _{v}^{2} \left(m_{h} \right)$ with size $n_{p} $ of halo groups. The flux function $\Pi _{kv} <0$ (inverse cascade) is normalized by ${Nm_{p} u_{0}^{2} /t_{0} } $ and computed from simulation results at two different redshifts $z_{1} $ and $z_{2} $ using Eq. \eqref{ZEqnNum299256} and dark matter flow dataset \citep{Xu:2022-Dark_matter-flow-dataset-part1}. A scale-independent flux function $\varepsilon _{kv}$ can be identified for propagation range with $m_{h} <m_{h}^{*} $.}
\label{fig:5}
\end{figure}

Just like the mass cascade in Eq. \eqref{ZEqnNum986318}, the flux functions of kinetic energy from halo velocity dispersion $\sigma _{h}^{2} $ and halo virial dispersion $\sigma _{v}^{2} $ are defined as (see Figs. \ref{fig:4} and \ref{fig:5}):
\begin{equation} 
\label{ZEqnNum299256} 
\begin{split}
&\Pi _{kh} =-\int _{m_{h} }^{\infty }\frac{\partial }{\partial t} \left[M_{h} \left(a\right)f_{M} \left(m,m_{h}^{*} \right)\right] \sigma _{h}^{2} \left(m,a\right)dm,\\ 
&\Pi _{kv} =-\int _{m_{h} }^{\infty }\frac{\partial }{\partial t} \left[M_{h} \left(a\right)f_{M} \left(m,m_{h}^{*} \right)\right] \sigma _{v}^{2} \left(m,a\right)dm.  
\end{split}
\end{equation} 
In mass propagation range, the energy flux functions are independent of mass scale $m_{h} $ \citep[see][Eqs. (27) and (48)]{Xu:2021-Inverse-and-direct-cascade-of-},
\begin{equation} 
\label{ZEqnNum126771} 
\begin{split}
&\varepsilon _{kh} \left(a\right)=\frac{3}{2} \Pi _{kh} =\frac{3}{2} \varepsilon _{m} \left\langle \sigma _{h}^{2} \right\rangle =-\frac{3}{4} M_{h} \left(a\right)H\left\langle \sigma _{h}^{2}\right\rangle,\\
&\varepsilon _{kv} \left(a\right)=\frac{3}{2} \Pi _{kv} =\frac{15}{2} \varepsilon _{m} \left\langle \sigma _{h}^{2} \right\rangle =-\frac{15}{4} M_{h} \left(a\right)H\left\langle \sigma _{v}^{2} \right\rangle,
\end{split}
\end{equation} 
where $\langle \sigma _{h}^{2} \rangle $ and $\langle \sigma _{v}^{2} \rangle $ are the average $\sigma _{h}^{2} $ and $\sigma _{v}^{2} $ for all halo particles in all halo groups. As expected, both flux functions are proportional to rate of mass transfer $\varepsilon _{m}$ in Eq. \eqref{eq:8}, since the inverse mass cascade facilitates the inverse energy cascade in dark matter flow. By contrast, the shape change of vortex (vortex stretching) is responsible for the energy cascade in hydrodynamic turbulence \citep{Xu:2021-Inverse-and-direct-cascade-of-}. 

Figures \ref{fig:4} and \ref{fig:5} present the variation of energy flux function $-\Pi _{kh} $ and $-\Pi _{kv} $ with the size $n_{p} $ of halo groups. It is computed from N-body simulation using Eq. \eqref{ZEqnNum299256} and dark matter flow dataset \citep{Xu:2022-Dark_matter-flow-dataset-part1}. Again, the propagation range with a scale-independent constant rate of energy transfer (Eq. \eqref{ZEqnNum126771}) and dissipation range can be identified in both figures. The total rate of kinetic energy cascade $\varepsilon _{u}$ (per unit mass) finally reads
\begin{equation} 
\label{ZEqnNum527910} 
\begin{split}
\varepsilon _{u} =\frac{\varepsilon _{kh} +\varepsilon _{kv} }{M} &=-\frac{9}{4} H\left\langle \sigma ^{2} \right\rangle \frac{M_{h} \left(a\right)}{M}\\
&\approx -\frac{9}{4} Hu^{2} =-\frac{3}{2} \frac{u^{2} }{t} =-4.6\times 10^{-7} \frac{m^{2} }{s^{3} },
\end{split}
\end{equation} 
where M is the total mass of all N-body system,  \textit{H} is the Hubble parameter, $t={2/\left(3H\right)} $ is the physical time and $t_{0} $ is present physical time. The mean dispersion $\left\langle \sigma _{h}^{2} \right\rangle =\left\langle \sigma _{v}^{2} \right\rangle \approx {\left\langle \sigma ^{2} \right\rangle /2} $, i.e. kinetic energy is equipartitioned between group temperature $\sigma _{h}^{2} $ and halo temperature $\sigma _{v}^{2} $. Here $\left\langle \sigma ^{2} \right\rangle $ is the one-dimensional velocity dispersion of all halo particles and $\left\langle \sigma ^{2} \right\rangle \approx u^{2} $, where $u^{2} $ is the velocity dispersion of all particles in entire N-body system. 

The rate of energy transfer $\varepsilon _{u} $ can also be obtained from the energy evolution of dark matter flow \citep[see][Eq. (41)]{Xu:2022-The-evolution-of-energy--momen}, which is consistent with Eq. \eqref{ZEqnNum527910}. The rate of energy cascade $\varepsilon _{u}$ is independent of both redshift and halo mass. It is a key parameter to determine dark matter particle mass and properties \citep[see][Eq. (19)]{Xu:2022-Postulating-dark-matter-partic}, the critical MOND acceleration \citep[see][Eq. (14)]{Xu:2022-The-origin-of-MOND-acceleratio}, dark matter halo properties \citep[see][Table 2] {Xu:2022-The-origin-of-MOND-acceleratio}, and the baryonic-to-halo mass ratio in this paper. 

\section{Baryonic-to-halo mass ratio from mass and energy cascade}
\label{sec:5}
The baryonic-to-halo mass ratio (BHMR) is the ratio of total baryonic mass ($m_{b} $, sum of cold gas and stellar mass) of a galaxy to the mass of dark matter halo ($m_{h} $) that galaxy resides in. The theory of mass/energy cascade can be used to analytically drive the baryonic-to-halo mass ratio. The SPARC (Spitzer Photometry \& Accurate Rotation Curves) data with $\mathrm{\sim}$175 late-type galaxies were used for developing the model \citep{Lelli:2019-The-baryonic-Tully-Fisher-rela}.

First, on the galactic scale, the baryonic Tully-Fisher relation (BTFR) \citep{Tully:1977-New-Method-of-Determining-Dist,McGaugh:2000-The-baryonic-Tully-Fisher-rela} is a natural result of the MOND theory, where the flat rotation velocity $v_{f}^{4} \propto m_{b} $. In "deep-MOND" regime, the Newtonian dynamics for a given point baryonic mass at a distance \textit{r} to halo center is modified to:\textbf{}
\begin{equation}
\label{ZEqnNum417694} 
\frac{Gm_{b} }{r^{2} } =\frac{\left({v_{f}^{2} /r} \right)^{2} }{a_{0} }  \Rightarrow  v_{f}^{4} =Gm_{b} a_{0} ,       
\end{equation} 
where $a_{0} $ is the critical MOND acceleration that might originate from the acceleration fluctuation in dark matter flow \citep{Xu:2022-The-origin-of-MOND-acceleratio}. This naturally leads to a flat rotation curve and BTFR in Eq. \eqref{ZEqnNum417694}.

Second, the relation between halo circular velocity $v_{cir} $ and halo virial size $r_{h} $ can be obtained using the virial theorem \citep{Xu:2021-Inverse-mass-cascade-halo-density},
\begin{equation} 
\label{ZEqnNum658276} 
v_{cir}^{2} =\frac{Gm_{h} }{r_{h} } =\frac{\Delta _{c} }{2} \left(Hr_{h} \right)^{2}  
\end{equation} 
and  
\begin{equation}
\begin{split}
&m_{h} =\frac{4}{3} \pi r_{h}^{3} \Delta _{c} \bar{\rho }_{0} a^{-3} \Rightarrow r_{h} \propto \left(m_{h} \right)^{{1/3} } a,\\
&v_{cir} =Hr_{h} \sqrt{{\Delta _{c}}/{2} } \propto \left(m_{h} \right)^{{1/3} } a^{{-1/2} }, 
\end{split}
\label{ZEqnNum246288}
\end{equation}

\noindent where Hubble constant $H_{0}^{2} ={8\pi G\bar{\rho }_{0} /3} $ and $\bar{\rho }_{0} \equiv \bar{\rho }\left(t=t_{0} \right)=\bar{\rho }\left(t\right)a^{3} $ is the mean matter density at $z=0$. Here $\Delta _{c} $ is the critical density ratio and $\Delta _{c} =18\pi ^{2} $ from the spherical collapse model or two-body collapse model (TBCM) \citep[see][Eq. (89)]{Xu:2021-A-non-radial-two-body-collapse}). 

Third, with the density ratio $\Delta _{c} =18\pi ^{2} $ in Eq. \eqref{ZEqnNum246288}, Eq. \eqref{ZEqnNum527910} for constant rate of energy cascade $\varepsilon _{u}$ can be rewritten as,
\begin{equation} 
\label{ZEqnNum693953} 
\varepsilon _{u} =\frac{3}{2} \frac{u^{2} }{t} =\frac{\left({3/2} \right)u^{2} }{{2\pi r_{h} /v_{cir} } } ,          
\end{equation} 
where ${t=2\pi r_{h} /v_{cir} } $ is the turnaround time, i.e. the period for DM particles to circling around the entire halo. Equation \eqref{ZEqnNum693953} describes the energy cascade in dark matter flow. The specific kinetic energy of dark matter particles is transferred across scales by $({3/2})u^{2}$ for every period of turnaround time \textit{t}. 

Now let us consider the flow of baryonic masses coupled to the flow of dark matter fluid that mimics a two-phase miscible flow. Due to the gravitational interaction between two "phases", the rate of energy cascade $\varepsilon_{u}$ is expected to be the same for the flow of both phases. In addition, the kinetic energy of baryonic mass has two contributions, i.e. from flat rotating speed $v_f^2$ and from the motion of halos, respectively. Therefore, just like Eq. \eqref{ZEqnNum693953}, $\varepsilon_{u}$ can be similarly related to the flat rotation speed $v_{f}$ and virial radius $r_{h}$ for the flow of baryons, but with different expressions for galaxies in small and large halos, respectively.

Small halos have a low peak height $\nu \equiv {\delta _{c} /\sigma _{\delta}} \left(m_{h} ,z\right)$ of density fluctuation (halos at their late stage with very slow mass accretion) \citep{Xu:2022-The-mean-flow--velocity-disper}. Halo velocity dispersion (group temperature defined in Eq. \eqref{ZEqnNum872828}) $\sigma _{h}^{2} $ is much greater than halo virial dispersion (halo temperature) $\sigma _{v}^{2} $ \citep[see][Fig. 2]{Xu:2021-Inverse-and-direct-cascade-of-}, i.e. the motion of halos is dominant over the random motion in halos. In small halos, the peculiar velocity of dark matter is of constant divergence for dark matter flow in small halos \citep{Xu:2022-The-statistical-theory-of-2nd,Xu:2022-The-mean-flow--velocity-disper}, while the proper velocity is incompressible. The baryonic masses are suspended in an incompressible dark matter fluid and two "phases" are intimately coupled together. The kinetic energy of baryonic mass in small halos (mostly from $\sigma _{h}^{2}$) is transferred by the same amount ($\sim u^{2} $) as that of dark matter in Eq. \eqref{ZEqnNum693953}, but with a different turnaround time of ${r_{h}/v_{f}}$ such that the rate of energy cascade reads (similar to Eq. \eqref{ZEqnNum693953}),
\begin{equation} 
\label{ZEqnNum521703} 
\varepsilon _{u} =-\beta _{f} \frac{u^{2} }{{r_{h} /v_{f} } } a^{q} ,          
\end{equation} 
where the constant $\beta _{f} $ and exponent $q$ are two parameters to be determined. With velocity dispersion of entire system $u^{2} \equiv u_{0}^{2} a^{{3/2} } $ (Eq. \eqref{ZEqnNum527910}), $r_{h} \propto a$ (Eq. \eqref{ZEqnNum246288}), and $v_{f} \propto a^{0}$ for virialized small halos, we would expect $q=-{1/2} $. Using Eqs. \eqref{ZEqnNum246288}, \eqref{ZEqnNum521703} and \eqref{ZEqnNum126771}, the halo circular velocity, halo virial size, and flat rotation speed for small halos should read,
\begin{equation} 
\label{ZEqnNum203805} 
\begin{split}
&v_{cir} =\frac{4}{9} \sqrt{\frac{\Delta _{c} }{2} } \beta _{f} v_{f} a^{q} \propto \left(m_{h} \right)^{{1/3} } a^{-{1/2} },\\
&r_{h} =\frac{4}{9} \beta _{f} v_{f} H^{-1} a^{q} \propto \left(m_{h} \right)^{{1/3} } a^{1},\\ 
&\textrm{and}\\
&v_{f} =\frac{9}{4\beta _{f} } \left(\frac{2}{\Delta _{c} } \right)^{\frac{1}{3} } \left(Gm_{h} H\right)^{{1/3} } a^{-q} \propto \left(m_{h} \right)^{{1/3} } a^{0}.
\end{split}
\end{equation} 

On the other hand, large halos with a high peak height $\nu $ are halos at their early stage with fast mass accretion and an almost constant halo concentration $c\approx 3.5$ \citep{Xu:2021-Inverse-mass-cascade-halo-density}. Halo velocity dispersion $\sigma _{h}^{2}$ is much smaller than halo virial dispersion $\sigma _{v}^{2} $ (Eq. \eqref{ZEqnNum872828}), i.e. large halos are much hotter. In addition, large halos are not incompressible such that two "phases" are not fully coupled. The kinetic energy transferred by the baryonic mass suspended in large halos is mostly from the rotational motion $v_f^2$. Therefore, just like the energy cascade in turbulence (Eq. \eqref{ZEqnNum629737}), the kinetic energy of baryonic mass is transferred by the amount proportional to $v_{f}^{2}$ for a turnaround period ${r_{h}/v_{f}}$ such that the rate of energy cascade reads (see Eq. \eqref{ZEqnNum644698}),
\begin{equation} 
\label{ZEqnNum886640} 
\varepsilon _{u} =-\alpha _{f} \frac{v_{f}^{2} }{{r_{h} /v_{f} } } a^{p} ,          
\end{equation} 
where the constant $\alpha _{f} $ and exponent $p$ are two parameters to be determined. Using Eqs. \eqref{ZEqnNum246288}, \eqref{ZEqnNum886640} and \eqref{ZEqnNum527910}, the halo circular velocity and halo size are proportional to $v_{f}^{3} $ for large halos,
\begin{equation} 
\label{ZEqnNum736837}
\begin{split}
&v_{cir} =\frac{4}{9} \sqrt{\frac{\Delta _{c} }{2} } \alpha _{f} \frac{v_{f}^{3} }{u^{2} } a^{p} \propto \left(m_{h} \right)^{{1/3} } a^{-{1/2} },\\
&r_{h} =\frac{4}{9} \alpha _{f} \frac{v_{f}^{3} }{Hu^{2} } a^{p} \propto \left(m_{h} \right)^{{1/3} } a^{1}\\ 
&\textrm{and}\\
&v_{f} =\left(\frac{3}{2\sqrt{\alpha _{f} } } \right)^{\frac{2}{3} } \left(\frac{2}{\Delta _{c} } \right)^{\frac{1}{9} } \left(Gm_{h} H\right)^{\frac{1}{9} } u^{\frac{2}{3} } a^{-\frac{p}{3} } \propto \left(m_{h} \right)^{\frac{1}{9} } a^{\frac{\left(1-p\right)}{3}}.
\end{split}
\end{equation} 

Equating quantities in Eqs. \eqref{ZEqnNum203805} and \eqref{ZEqnNum736837} leads to a critical flat rotation velocity $v_{fc} $ or critical circular velocity $v_{cc} $, a critical halo size $r_{hc}$, or a critical halo mass $m_{hc}$ to delineate two different regimes,
\begin{equation} 
\label{ZEqnNum832262}
\begin{split}
&v_{fc} =ua^{{\left(q-p\right)/2} } \sqrt{{\beta _{f} /\alpha _{f} } } ,\\
&v_{cc} =\frac{4}{9} \sqrt{\frac{\Delta _{c} }{2} } \sqrt{\frac{\beta _{f}^{3} }{\alpha _{f} } } ua^{{\left(3q-p\right)/2} },\\
&r_{hc} =\frac{4}{9} a^{{\left(3q-p\right)/2} } uH^{-1} \beta _{f} \sqrt{{\beta _{f} /\alpha _{f} } },\\ 
&m_{hc} =\left(\frac{\beta _{f}^{3} }{\alpha _{f} } \right)^{\frac{3}{2} } \left(\frac{8\Delta _{c}}{81} \right)\left(\frac{u^5}{G\varepsilon_{u}} \right)a^{\frac{3}{2} \left(3q-p\right)}.
\end{split}
\end{equation} 
where $m_{hc}$ is determined by the three quantities, i.e. the velocity dispersion $u$, gravitational constant $G$, and constant rate of energy cascade $\varepsilon_u$ \citep[also see][Table 2 for a similar result derived from a simple dimensional analysis on large scale]{Xu:2022-The-origin-of-MOND-acceleratio}. 

\begin{figure}
\includegraphics*[width=\columnwidth]{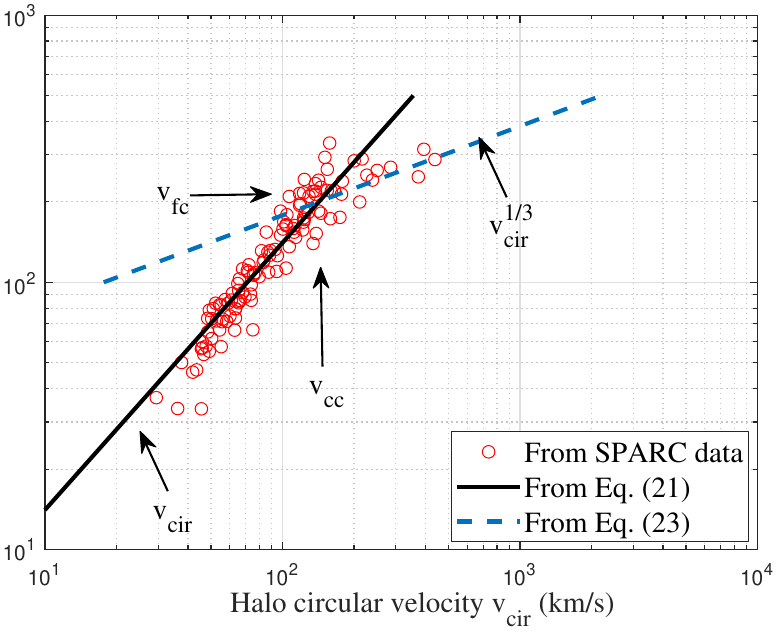}
\caption{The variation of flat rotation velocity $v_{f}$ (km/s) with halo circular velocity $v_{cir} $. For small halos, $v_{f} \propto v_{cir} $ (black solid) with $\Delta _{c} =200$ and $\beta _{f} \approx 0.16$ in Eq. \eqref{ZEqnNum203805}. For large halos, $v_{f} \propto \left(v_{cir} \right)^{{1/3} } $ (blue dash) with $\Delta _{c} =200$ and $\alpha _{f} \approx 0.5$ in Eq. \eqref{ZEqnNum736837}. The critical flat rotation velocity $v_{fc} \approx 200{km/s} $ and $v_{fc} \approx 143{km/s} $ at \textit{z}=0.}
\label{fig:6}
\end{figure}

\begin{figure}
\includegraphics*[width=\columnwidth]{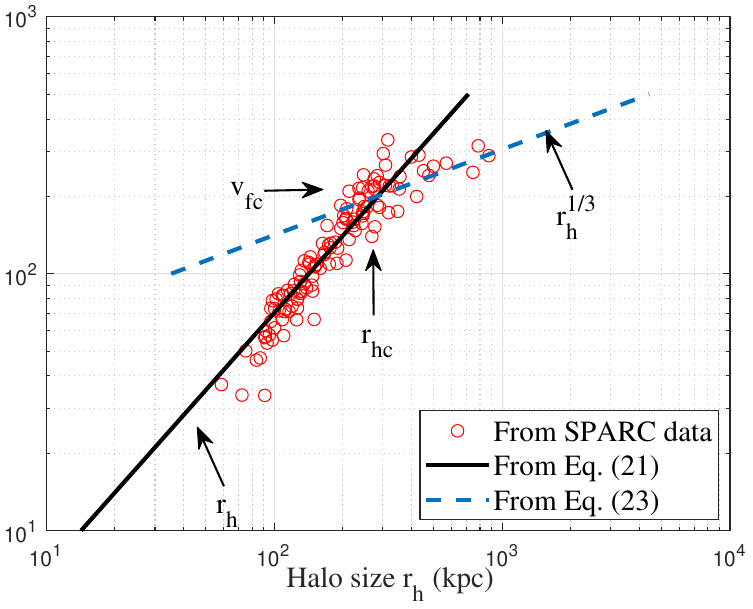}
\caption{The variation of $v_{f}$ (km/s) with halo size $r_{h}$. For small halos, $v_{f} \propto r_{h} $ (black solid) with $\beta _{f} \approx 0.16$ in Eq. \eqref{ZEqnNum203805}. For large halos, $v_{f} \propto \left(r_{h} \right)^{{1/3} } $ (blue dash) with $\alpha _{f} \approx 0.5$ in Eq. \eqref{ZEqnNum736837}. Critical size $r_{hc} \approx 285kpc$ at \textit{z}=0.}
\label{fig:7}
\end{figure}

\begin{figure}
\includegraphics*[width=\columnwidth]{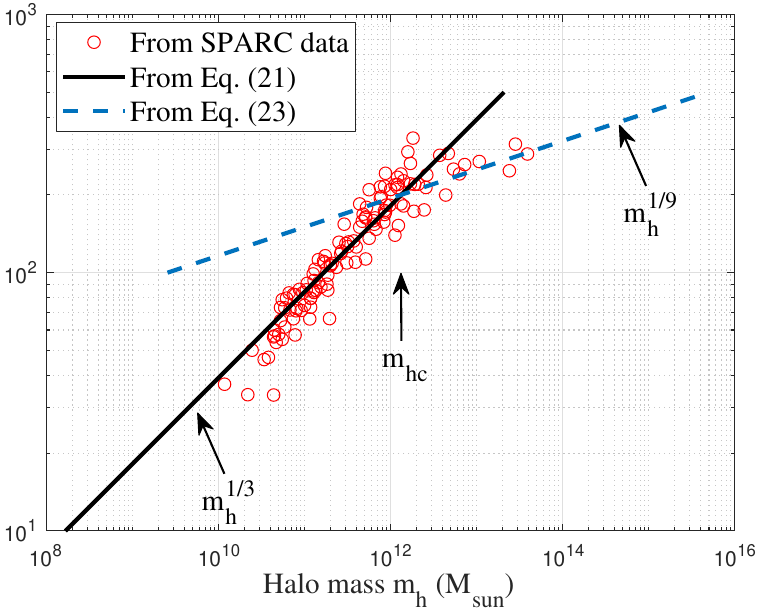}
\caption{The variation of $v_{f}$ (km/s) with halo mass $m_{h}$. For small halos, $v_{f} \propto \left(m_{h} \right)^{{1/3} } $ (black solid) with $\Delta _{c} =200$ and $\beta _{f} \approx 0.16$ in Eq. \eqref{ZEqnNum203805}. For large halos, $v_{f} \propto \left(m_{h} \right)^{{1/9} } $ (blue dash) with $\Delta _{c} =200$ and $\alpha _{f} \approx 0.5$ (Eq. \eqref{ZEqnNum736837}). The critical halo mass $m_{hc} \approx 1.33\times 10^{12} M_{\odot} $ at \textit{z}=0.}
\label{fig:8}
\end{figure}

With parameters in Table \ref{tab:2}, Figures \ref{fig:6}, \ref{fig:7}, and \ref{fig:8} plot the variation of flat rotation velocity $v_{f} $ with halo circular velocity $v_{cir} $, halo size $r_{h} $, and halo mass $m_{h} $ from SPARC data \citep{Lelli:2016-SPARC-Mass-Models-for-175-Disk-Galaxies,Lelli:2019-The-baryonic-Tully-Fisher-rela}, along with two regimes from Eqs. \eqref{ZEqnNum203805} and \eqref{ZEqnNum736837} for small and large halos. At \textit{z}=0, the critical rotation velocity $v_{fc} \approx 200{km/s} $, circular velocity $v_{cc} \approx 143{km/s} $, and halo size $r_{hc} \approx 285kpc$. The critical halo mass $m_{hc} \approx 1.33\times 10^{12} M_{\odot} $ is consistent with the halo mass with the greatest stellar-to-halo mass ratio \citep{Moster:2013-Galactic-star-formation-and-ac,Moster:2010-Constraints-on-the-Relationshi,Girelli:2020-The-stellar-to-halo-mass-relat}. The scaling from SPARC data are in agreement with predictions in Eqs. \eqref{ZEqnNum203805} and \eqref{ZEqnNum736837}. 

\begin{figure}
\includegraphics*[width=\columnwidth]{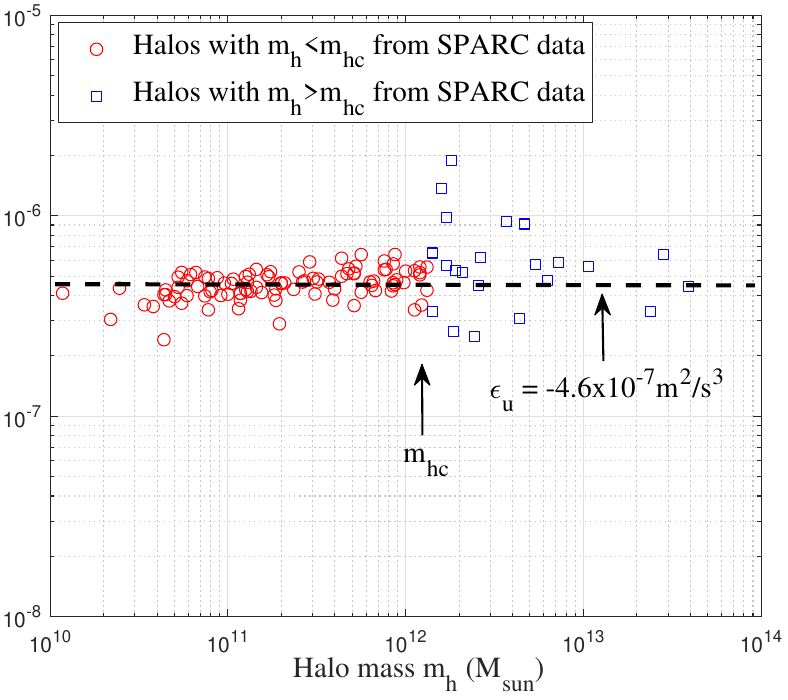}
\caption{The rate of energy cascade $\epsilon_u$ (\textrm{$m^2/s^3$}) for 153 halos from SPARC data using Eq. \eqref{ZEqnNum521703} for halos $m_{h} <m_{hc}$ (red circle) with $\beta _{f} \approx 0.16$, and using Eq. \eqref{ZEqnNum886640}  for large halos $m_{h} >m_{hc} $ (blue square) with $\alpha _{f} \approx 0.5$. Critical halo mass $m_{hc} \approx 1.33\times 10^{12} M_{\odot}$. Halos have different rate of energy cascade (spatial intermittency). The average $\varepsilon _{u} \approx 4.6\times 10^{-7} {m^{2} /s^{3}}$.}
\label{fig:9}
\end{figure}

In fact, the rate of energy cascade in dark matter flow can be of an intermittent nature in space. The overall rate of energy cascade for entire system ($\varepsilon _{u} $) is a globally averaged quantity. For different halos at different locations, the rate of energy cascade of individual halos can be different, i.e. there can be a distribution of the rate of energy cascade for all halos. Just like the intermittency in hydrodynamic turbulence, the intermittency in dark matter flow and its effects on flow dynamics are important topics to explore in the future. Figure \ref{fig:9} plots the rate of energy cascade for small ($<m_{hc} $) and large halos ($>m_{hc} $) from SPARC data using Eqs. \eqref{ZEqnNum521703} and \eqref{ZEqnNum886640}.

Finally, combining the baryonic Tully-Fisher relation in Eq. \eqref{ZEqnNum417694} and the rate of energy cascade for small halos $m_{h} <m_{hc} $ in Eq. \eqref{ZEqnNum521703} , along with the expression of halo size $r_{h} $ from Eq. \eqref{ZEqnNum246288}, the baryonic to halo mass relation for small halos can be analytically derived,
\[m_{b} =\left(M_{c1} \right)^{-{1/3} } \left(m_{h} \right)^{{4/3} },\] 
where the characteristic mass $M_{c1} $ reads  
\begin{equation} 
\label{ZEqnNum388435} 
\begin{split}
M_{c1} \left(a\right)&=\left(\beta _{f} a^{q} \right)^{12} \left(\frac{\Delta _{c} }{2} \right)^{4} \left(\frac{u^{24} H^{8} a_{0}^{3} }{G\varepsilon _{u}^{12} } \right)\\
&=\left(\frac{2}{3} \right)^{16} \left(\beta _{f} a^{q} \right)^{12} \left(\frac{\Delta _{c} }{2} \right)^{7} \left(\frac{u^{5} }{G\varepsilon _{u} } \right).
\end{split}
\end{equation} 
The baryonic-to-halo mass ratio simply reads, 
\begin{equation} 
\label{eq:26} 
{m_{b} /m_{h} } =\left({m_{h} /M_{c1} } \right)^{{1/3} } =\left({m_{b} /M_{c1} } \right)^{{1/4} } ,        
\end{equation} 
where the baryon fraction $m_{b} /m_{h}\propto m_{b}^{{1/4} } $ that agrees well with literature \citep{Chan:2019-A-universal-constant-for-dark-}.

Similarly, combining the baryonic Tully-Fisher Eq. \eqref{ZEqnNum417694},  the rate of energy cascade for large halos $m_{h} >m_{hc} $ in Eq. \eqref{ZEqnNum886640}, and Eq. \eqref{ZEqnNum246288}, the baryonic mass is related to halo mass for large halos,
\[m_{b} =\left(M_{c2} \right)^{{5/9} } \left(m_{h} \right)^{{4/9} },\] 
where the characteristic mass $M_{c2} $ reads  
\begin{equation} 
\label{ZEqnNum532889} 
\begin{split}
M_{c2} \left(a\right)&=\left[\left(\frac{1}{\alpha _{f} a^{p} } \right)^{12} \left(\frac{2}{\Delta _{c} } \right)^{4} \frac{\varepsilon _{u}^{12} }{G^{5} H^{8} a_{0}^{9} } \right]^{{1/5} }\\ 
&=\left(\frac{2}{3} \right)^{-\frac{16}{5} } \left(\alpha _{f} a^{p} \right)^{-\frac{12}{5} } \left(\frac{2}{\Delta _{c} } \right)^{\frac{13}{5} } \left(\frac{u^{5} }{G\varepsilon _{u} } \right).
\end{split}
\end{equation} 
The baryonic-to-halo mass ratio ${m_{b} /m_{h} } \propto \left(m_{h} \right)^{{-5/9} }$ in large halos also agrees well with relevant study \citep{Moster:2010-Constraints-on-the-Relationshi}. 

There exists a critical halo mass $m_{hc} $ where the baryonic mass $m_{b} $ of small and large halos in Eqs. \eqref{ZEqnNum388435} and \eqref{ZEqnNum532889} are equal (the same critical halo mass as Eq. \eqref{ZEqnNum832262}),
\begin{equation} 
\label{eq:28} 
m_{hc} =\left[\left(M_{c1} \right)^{3} \left(M_{c2} \right)^{5} \right]^{{1/8}}.     
\end{equation} 
With Eq. \eqref{ZEqnNum527910}, the critical halo mass can be rewritten as,
\begin{equation} 
\label{eq:29} 
m_{hc} =\frac{16}{81} \left(\frac{\beta _{f}^{3} }{\alpha _{f} } \right)^{{3/2} } \left(\frac{\Delta _{c} }{2} \right)\left(\frac{u^{5} }{G\varepsilon _{u}^{} } \right)a^{\frac{3}{2} \left(3q-p\right)}.  
\end{equation} 
The critical baryonic mass $m_{bc}$ in halos with a critical mass $m_{hc}$ is 
\begin{equation} 
\label{eq:30} 
m_{bc} =\left[\left(M_{c1} \right)\left(M_{c2} \right)^{5} \right]^{{1/6} } =\left(\frac{\beta _{f} }{\alpha _{f} } \right)^{2} \left(\frac{u^{4} }{Ga_{0} } \right)a^{2\left(q-p\right)} .      
\end{equation} 
With the critical acceleration $a_0$ related to the rate of energy cascade as $a_{0} \left(a\right)=-{\Delta _{c} \varepsilon _{u} /\left(2u\right)} $\citep{Xu:2022-The-origin-of-MOND-acceleratio}, the critical baryonic mass is
\begin{equation} 
\label{eq:31} 
m_{bc} =\frac{2}{\Delta _{c} } \left(\frac{\beta _{f} }{\alpha _{f} } \right)^{2} \left(\frac{u^{5} }{G\varepsilon _{u} } \right)a^{2\left(q-p\right)} .        
\end{equation} 

Since $m_{b} \propto \left(m_{h} \right)^{{4/3} } $ for small halos and $m_{b} \propto \left(m_{h} \right)^{{4/9} } $ for large halos (see Table \ref{tab:2} and Fig. \ref{fig:11}), there exist a maximum baryonic-to-halo mass ratio for halos with critical mass $m_{hc}$
\begin{equation} 
\label{ZEqnNum272715} 
A\left(z\right)\equiv \frac{m_{bc} }{m_{hc} } =\left(\frac{M_{c2} }{M_{c1} } \right)^{{5/24} } =\frac{{2/\Delta _{c} } }{\left(\alpha _{f} \right)^{{1/2} } \left(\beta _{f} \right)^{{5/2} } } \gamma _{f} a^{-{\frac{5q+p}{2}} } ,      
\end{equation} 
where $\gamma _{f} $ is a dimensionless redshift-independent constant that sets the scale of baryonic-to-halo mass ratio (see Eq. \eqref{ZEqnNum527910} and relation $\varepsilon _{u} =-{a_{0}u/(3\pi) ^{2}}$ \citep{Xu:2022-The-origin-of-MOND-acceleratio}),
\begin{equation} 
\label{ZEqnNum850635} 
\gamma _{f} =\frac{\varepsilon _{u}^{3} }{H^{2} u^{5} a_{0} } =\frac{81}{8\Delta _{c} } \approx \left(\frac{3}{4\pi } \right)^{2} =0.057.       
\end{equation} 
The maximum baryonic-to-halo mass ratio is $A\left(z=0\right)\approx 0.076$ with all relevant parameters listed in Table \ref{tab:2}. 

\section{Redshift evolution of fraction of baryons}
\label{sec:6}
A rough estimate for the fraction of total baryonic mass in all halos can be made here. It is well known that the overall cosmic baryonic-to-DM mass ratio (including both halos and out-of-halo) is $\mathrm{\sim}$18.8\% in standard $\Lambda$CDM model. By dividing entire system into the halo subsystem (masses in all halos) and out-of-halo subsystem (masses not belong to any halos), the baryonic-to-DM mass ratio in out-of-halo subsystem $A_{boh}(z)$ reads 
\begin{equation} 
\label{ZEqnNum269995} 
A_{boh} \left(z\right)=\frac{0.188-A_{dh} \left(z\right)A_{bh} \left(z\right)}{1-A_{dh} \left(z\right)} ,        
\end{equation} 
where $A_{dh} \left(z\right)$ is the fraction of total cosmic dark matter that resides in halo subsystem and $A_{bh} \left(z\right)$ is the average baryonic-to-halo mass ratio in halo subsystem for all halos of different size. 

At \textit{z}=0, due to continuous inverse mass cascade \citep{Xu:2021-Inverse-mass-cascade-mass-function}, $\mathrm{\sim}$60\% of dark matter forms halos and $\mathrm{\sim}$40\% dark matter does not belong to any halo (out-of-halo), and $A_{dh} \left(z\right)\approx 0.6a^{{1/2} } $ with total halo mass $M_{h} \left(z\right)\propto a^{{1/2}}$ \citep[see][Table 2]{Xu:2021-Inverse-mass-cascade-mass-function}. First, for simplicity, let us assume the baryonic-to-halo mass ratio in halo subsystem is just the maximum ratio $A\left(z\right)$, i.e. the average ratio $A_{bh} \left(z\right)\equiv A\left(z\right)\approx 0.076$, the minimum baryonic-to-DM mass ratio in out-of-halo system $A_{boh} \left(z\right)$ is $\mathrm{\sim}$35.6\% from Eq. \eqref{ZEqnNum269995} and a maximum fraction ${0.6A_{bh} \left(z\right)/0.188\approx 24.3\% } $ of total baryonic mass are in halo subsystem (all halos).

\begin{figure}
\includegraphics*[width=\columnwidth]{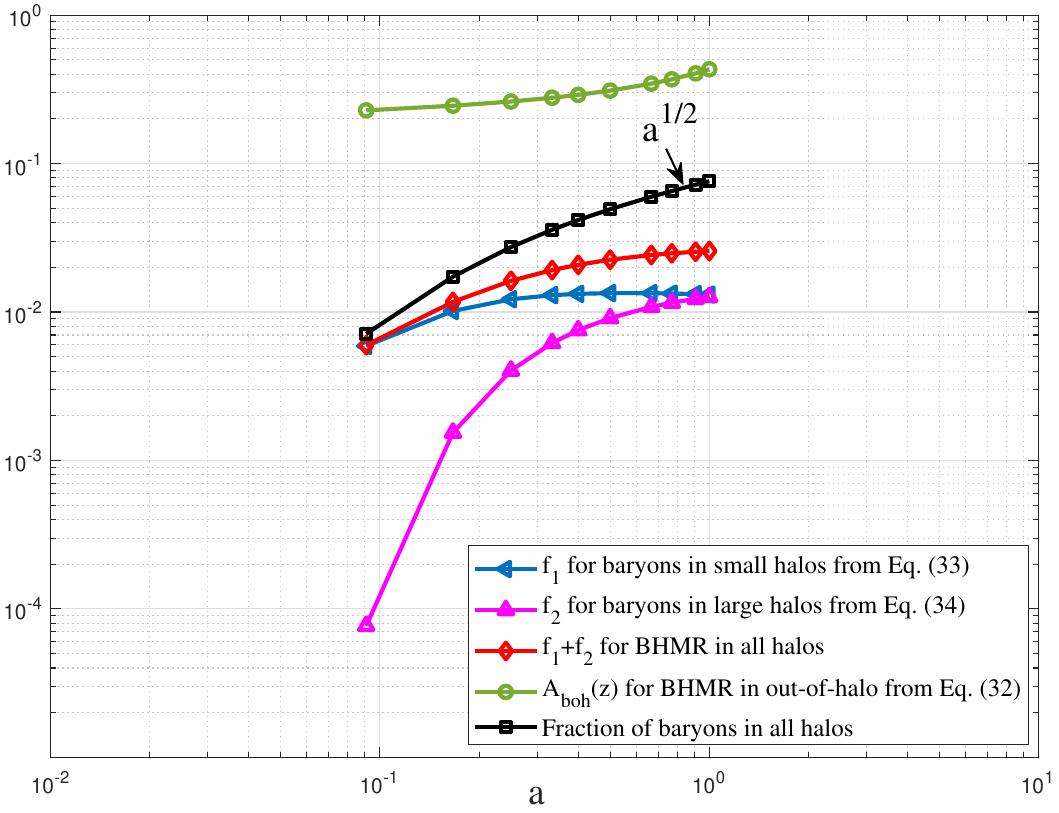}
\caption{The time variation of the baryonic-to-halo mass ratio (BHMR) in all halos (red diamond), the baryonic-to-DM ratio in out-of-halo (green circle), and the fraction of baryonic mass in halos (black square) that increases approximately $\propto a^{{1/2} } $. At \textit{z}=0, the average BHMR in all halos is $\mathrm{\sim}$0.024, the baryonic-to-DM ratio in out-of-halo is $\mathrm{\sim}$0.434, and the $\mathrm{\sim}$7.6\% of total baryons are in halos.}
\label{fig:10}
\end{figure}

More accurate estimate can be derived using the halo mass function (e.g. Eq. \eqref{ZEqnNum889111}) and Eqs. \eqref{ZEqnNum388435} and \eqref{ZEqnNum532889} to find the average baryonic-to-halo mass ratio $A_{bh} \left(z\right)$ for all halos. The baryonic-to-halo mass ratio can be separated into two contributions $f_1$ from small and $f_2$ from large halos, respectively,
\begin{equation} 
\label{ZEqnNum855755}
\begin{split}
f_{1}&=\int _{0}^{\nu _{c} }f_{D\lambda }  \left(\nu \right)\left(M_{c1} \right)^{-{1/3} } \left(\nu ^{{3/2} } m_{h}^{*} \right)^{{1/3} } d\nu \\
&=\left(\frac{m_{h}^{*} }{M_{c1} } \right)^{{1/3} } \frac{\left(2\sqrt{\eta _{0} } \right)}{\Gamma \left({q_{0} /2} \right)}\left[\Gamma \left(\frac{1+q_{0} }{2} \right)-\Gamma \left(\frac{1+q_{0} }{2} ,\frac{\nu _{c} }{4\eta _{0} } \right)\right],
\end{split}
\end{equation} 
\begin{equation} 
\label{ZEqnNum227267} 
\begin{split}
f_{2}&=\int _{\nu _{c} }^{\infty }f_{D\lambda }  \left(\nu \right)\left(M_{c2} \right)^{{5/9} } \left(\nu ^{{3/2} } m_{h}^{*} \right)^{{-5/9} } d\nu \\
&=\left(\frac{M_{c2} }{m_{h}^{*} } \right)^{{5/9} } \frac{\left(2\sqrt{\eta _{0} } \right)^{-q_{0} } }{\Gamma \left({q_{0} /2} \right)}\left(\nu _{c} \right)^{\frac{q_{0} }{2} -\frac{5}{6} } Ei\left(\frac{11}{6} -\frac{q_{0} }{2} ,\frac{\nu _{c} }{4\eta _{0} } \right). 
\end{split}
\end{equation} 
where dimensionless variable $\nu =({m_{h} /m_{h}^{*} })^{{2/3} } $with a critical value of $\nu _{c} =({m_{hc} /m_{h}^{*} })^{{2/3} } $. Here $\Gamma(x,y)$ is an upper incomplete gamma function and $Ei(x,y)$ is a two-parameter exponential integral function. The following identity of integration is used,
\begin{equation} 
\label{eq:37} 
\int _{0}^{\nu _{c} }\nu ^{s} \exp \left(-\frac{\nu }{4\eta _{0} } \right)d\nu  =\left(4\eta _{0} \right)^{1+s} \Gamma \left(1+s,\frac{\nu _{c} }{4\eta _{0} } \right),      
\end{equation} 
\begin{equation} 
\label{eq:38} 
\int _{\nu _{c} }^{\infty }\nu ^{s} \exp \left(-\frac{\nu }{4\eta _{0} } \right)d\nu  =\left(\nu _{c} \right)^{1+s} Ei\left(-s,\frac{\nu _{c} }{4\eta _{0} } \right).       
\end{equation} 

With relevant parameters listed in Table \ref{tab:2}, Figure \ref{fig:10} presents the time variation of baryonic-to-halo mass ratio (BHMR) in halo subsystem (Eqs. \eqref{ZEqnNum855755} and \eqref{ZEqnNum227267}) that reaches a constant value of $A_{bh}(z=0)\approx 0.024$ and the baryonic-to-DM ratio in out-of-halo subsystem ($\mathrm{\sim}$43.4\% from Eq. \eqref{ZEqnNum269995}). Since total mass of dark matter in all halos $M_{h}(z)\propto a^{{1/2}}$, the total baryonic mass in all halos should be $0.024M_h(z)\propto a^{1/2}$.

At \textit{z}=0, with about 60\% of DM in halos, the fraction of total baryons residing in halo subsystem (${0.6A_{bh} \left(z\right)/0.188} $) is estimated to be $\mathrm{\sim}$7.6\%. This theoretical estimation agrees well with that from various large astronomical surveys \citep{Read:2005-The-baryonic-mass-function-of-}, where baryons in galaxy $\Omega _{b,gal} \approx 0.0035$ and total baryons $\Omega _{b} \approx 0.046$ such that the fraction $\Omega _{b,gal}/\Omega _{b}=0.076$. That fraction increases with time as $\propto a^{{1/2}}$ (Fig. \ref{fig:10}), while majority of baryons ($92.4\%$) are not residing in any halos. Consequently, we have the mass fraction for baryons and dark matter in both halos and out-of-halos (subscript "oh"): $\Omega_{b,gal}=0.0035$, $\Omega_{b,oh}=0.0425$, $\Omega_{dm,halo}=0.147$, and $\Omega_{dm,oh}=0.098$. The baryonic mass function of galaxies can be subsequently derived with halo mass function and baryonic-to-halo mass relation established.

\begin{figure}
\includegraphics*[width=\columnwidth]{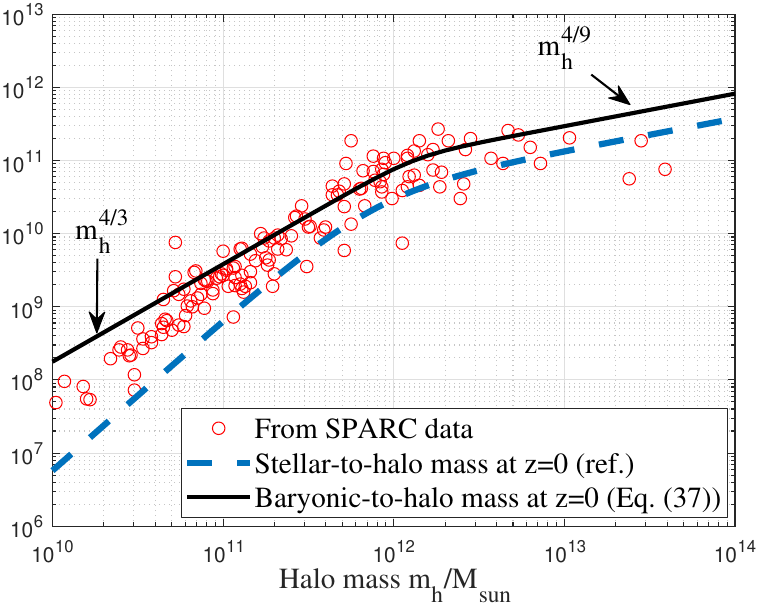}
\caption{The variation of baryonic mass $m_b$ ($M_{\odot}$)with halo mass for 153 halos from SPARC data. Model from Eq. \eqref{ZEqnNum395744} is plotted, where $m_{b} \propto \left(m_{h} \right)^{{4/3} } $ for small halos $m_{h} <m_{hc} $. For large halos $m_{h} >m_{hc} $, the baryonic mass $m_{b} \propto \left(m_{h} \right)^{{4/9} } $and agrees with the stellar-to-halo mass ratio required to reproduce the stellar mass function \citep[see][]{Moster:2010-Constraints-on-the-Relationshi} (blue dash line).}
\label{fig:11}
\end{figure}

Note that combining results for small and large halos together (Eqs. \eqref{ZEqnNum388435} and \eqref{ZEqnNum532889}), the baryonic-to-halo mass ratio can be conveniently modelled by a double-power-law that is similar to the empirical models for stellar-to-halo mass ratio from astronomical observations \citep{Moster:2010-Constraints-on-the-Relationshi,Girelli:2020-The-stellar-to-halo-mass-relat},\textbf{}
\begin{equation} 
\label{ZEqnNum395744} 
\frac{m_{b} }{m_{h} } =2^{\frac{1}{m} } A\left(z\right)\left[\left(\frac{m_{h} }{m_{hc} \left(z\right)} \right)^{-\frac{m}{3} } +\left(\frac{m_{h} }{m_{hc} \left(z\right)} \right)^{\frac{5m}{9} } \right]^{-\frac{1}{m} } ,      
\end{equation} 
where parameter $m$ adjusts the sharpness of the transition between two regimes. The redshift dependence of maximum stellar-to-halo mass ratio roughly follows $\propto 0.046a^{0.38} $ \citep{Girelli:2020-The-stellar-to-halo-mass-relat}. By assuming the baryonic-to-halo mass ratio $A\left(z\right)$ in Eq. \eqref{ZEqnNum272715} follows the same scaling as the stellar-to-halo mass ratio, we should have $p={7/4} $ for $q=-{1/2} $. Another reasonable option can be $p={3/2}$ for $q=-{1/2} $, but will require more observation data to confirm. For halos smaller than $m_{hc}$, the accretion of dark matter is much slower than baryons such that the baryonic-to-halo mass ratio increases with time. For halos greater than $m_{hc}$, the accretion of dark matter is still dominant over baryons such that the baryonic-to-halo mass ratio remains constant or slowly decreases with time. Therefore, the critical mass $m_{hc}$ with the greatest baryonic-to-halo mass ratio $A(z)$ should decrease with time. 

Figure \ref{fig:11} plots the variation of baryonic mass with halo mass for 153 halos from SPARC data. The derived model in Eq. \eqref{ZEqnNum395744} is also plotted (solid black) for comparison. The stellar-to-halo mass ratio required to reproduce the observed stellar mass function is also presented in the same figure (dash blue) \citep{Moster:2010-Constraints-on-the-Relationshi}. For large halos, the stellar mass follows the same power-law as baryonic mass derived in Eq. \eqref{ZEqnNum395744}. 

All current analysis is based on the baryonic Tully-Fisher relation with an exact exponent 4 in Eq. \eqref{ZEqnNum417694}. However, similar analysis can be extended to a different scaling of baryonic Tully-Fisher relation with a slightly different exponent.

\begin{table}
\caption{Parameters for deriving baryonic-to-halo mass ratio}
\begin{tabular}{|p{0.09in}|p{0.75in}|p{0.09in}|p{0.18in}|p{0.18in}|p{1.1in}|} \hline 
$\Delta _{c} $ & $200$ & $p$ & ${7/4} $ & $M_{c1} $ & $3.01\times 10^{15} a^{-{9/4} } M_{\odot} $ \\ \hline 
$\varepsilon _{u} $ & $4.6\times 10^{-7} {m^{2} /s^{3} } $ &  $q$ & $-{1/2} $ &  $M_{c2} $ & $1.29\times 10^{10} a^{-{9/20} } M_{\odot} $ \\ \hline 
$H_{0} $ & $1.62\times 10^{-18} {1/s} $ &  $\alpha _{f} $ & $0.5$ &  $m_{hc} $ & $1.33\times 10^{12} a^{-{9/8} } M_{\odot} $ \\ \hline 
$u_{0} $ & $354.61{km/s} $ & $\beta _{f} $ & $0.16$ &  $m_{bc} $ & $1.01\times 10^{11} a^{-{3/4} } M_{\odot} $ \\ \hline 
$a_{0} $ & $1.2\times 10^{-10} {m/s^{2} } $ & $m$ & $4$ & $A\left(z\right)$ & $0.0761a^{{3/8} } $ \\ \hline 
$\eta _{0} $ & $0.76$ & $q_{0} $ & $0.556$ & $m_{h}^{*} $ & $4\times 10^{13} a^{{3/2} } M_{\odot} $ \\ \hline 
\end{tabular}
\label{tab:2}
\end{table}

\section{Conclusions}
\label{sec:7}
Main focus of this paper is to apply the unique properties of self-gravitating collisionless dark matter flow (SG-CFD) to derive the baryonic-to-halo mass relation for halos of different size. In dark matter flow, the long-range interaction requires a broad size of halos to be formed to maximize system entropy. Halos facilitate an inverse mass cascade from small to large mass scales and an inverse (kinetic) energy cascade with a constant rate of energy transfer $\varepsilon _{u} \approx -4.6\times 10^{-7} {m^{2} /s^{3} } $. The mass/energy cascade represents an intermediate statistically steady state of dark matter flow. Unlike the hydrodynamic turbulence that is incompressible on all scales, dark matter flow also exhibits scale-dependent flow behaviors on different scales. 

The baryonic fluid in halos interacts with dark matter fluid through gravity and should share the same rate of energy cascade as that of dark matter flow. The kinetic energy of baryons is dominated by the motion of halos for small halos, and dominated by the rotational motion of barons for large halos. Combining the energy cascade from the motion of baryons and the baryonic Tully-Fisher relation, the baryonic-to-halo mass ratio can be analytically derived with a maximum ratio $\mathrm{\sim}$0.076 for halos with a critical mass of $\sim 10^{12} M_{\odot}$ at \textit{z}=0. That ratio is much smaller for both smaller and larger halos. 

For galaxies/halos with a flat rotation velocity $v_{f} $, baryonic mass $m_{b} $, halo mass $m_{h} $, and halo size $r_{h} $, two regimes can be clearly identified: for small incompressible halos $m_{h} <m_{hc} $, we have $\varepsilon _{u} \propto {v_{f} /r_{h} } $, $v_{f} \propto r_{h} $ and $m_{b} \propto \left(m_{h} \right)^{{4/3} } $. While for large halos $m_{h} >m_{hc} $, we have $\varepsilon _{u} \propto {v_{f}^{3} /r_{h} } $, $v_{f} \propto r_{h}^{{1/3}}$ and $m_{b} \propto \left(m_{h} \right)^{{4/9} } $. The spatial intermittency of energy cascade might be demonstrated by the variation of $\varepsilon_u$ between different halos. Its effects on the dynamics and evolution of individual halos should be explored in the future.

Combined with the double-$\lambdaup$ halo mass function, the average BHMR ratio in all halos ($\mathrm{\sim}$0.024 at z=0) can be analytically derived, along with its redshift evolution. The fraction of baryons in all galaxies is only $\mathrm{\sim}$7.6\% at z=0 and increases with time $\propto t^{{1/3}}$. Majority of baryons ($92.4\%$) are not residing in any halos. 
 

\section*{Data Availability}
Two datasets underlying this article, i.e. a halo-based and correlation-based statistics of dark matter flow, are available on Zenodo \citep{Xu:2022-Dark_matter-flow-dataset-part1,Xu:2022-Dark_matter-flow-dataset-part2}, along with the accompanying presentation slides "A comparative study of dark matter flow \& hydrodynamic turbulence and its applications" \citep{Xu:2022-Dark_matter-flow-and-hydrodynamic-turbulence-presentation}. All data files are also available on GitHub \citep{Xu:Dark_matter_flow_dataset_2022_all_files}.

\bibliographystyle{mnras}
\bibliography{Papers}

\begin{thebibliography}{}
\makeatletter
\relax
\def\mn@urlcharsother{\let\do\@makeother \do\$\do\&\do\#\do\^\do\_\do\%\do\~}
\def\mn@doi{\begingroup\mn@urlcharsother \@ifnextchar [ {\mn@doi@}
  {\mn@doi@[]}}
\def\mn@doi@[#1]#2{\def\@tempa{#1}\ifx\@tempa\@empty \href
  {http://dx.doi.org/#2} {doi:#2}\else \href {http://dx.doi.org/#2} {#1}\fi
  \endgroup}
\def\mn@eprint#1#2{\mn@eprint@#1:#2::\@nil}
\def\mn@eprint@arXiv#1{\href {http://arxiv.org/abs/#1} {{\tt arXiv:#1}}}
\def\mn@eprint@dblp#1{\href {http://dblp.uni-trier.de/rec/bibtex/#1.xml}
  {dblp:#1}}
\def\mn@eprint@#1:#2:#3:#4\@nil{\def\@tempa {#1}\def\@tempb {#2}\def\@tempc
  {#3}\ifx \@tempc \@empty \let \@tempc \@tempb \let \@tempb \@tempa \fi \ifx
  \@tempb \@empty \def\@tempb {arXiv}\fi \@ifundefined
  {mn@eprint@\@tempb}{\@tempb:\@tempc}{\expandafter \expandafter \csname
  mn@eprint@\@tempb\endcsname \expandafter{\@tempc}}}

\bibitem[\protect\citeauthoryear{Batchelor}{Batchelor}{1953}]{Batchelor:1953-The-Theory-of-Homogeneous-Turb}
Batchelor G.~K.,  1953, The Theory of Homogeneous Turbulence.
Cambridge University Press, Cambridge, UK

\bibitem[\protect\citeauthoryear{Behroozi, Conroy  \& Wechsler}{Behroozi
  et~al.}{2010}]{Behroozi:2010-A-COMPREHENSIVE-ANALYSIS-OF-UN}
Behroozi P.~S.,  Conroy C.,   Wechsler R.~H.,  2010, \mn@doi [Astrophysical
  Journal] {10.1088/0004-637x/717/1/379}, 717, 379

\bibitem[\protect\citeauthoryear{Bower}{Bower}{1991}]{Bower:1991-The-Evolution-of-Groups-of-Gal}
Bower R.~G.,  1991, \mn@doi [Monthly Notices of the Royal Astronomical Society]
  {10.1093/mnras/248.2.332}, 248, 332

\bibitem[\protect\citeauthoryear{Chan}{Chan}{2019}]{Chan:2019-A-universal-constant-for-dark-}
Chan M.~H.,  2019, \mn@doi [Scientific Reports] {10.1038/s41598-019-39717-x},
  9, 3570

\bibitem[\protect\citeauthoryear{Colberg, White, Jenkins  \& Pearce}{Colberg
  et~al.}{1999}]{Colberg:1999-Linking-cluster-formation-to-l}
Colberg J.~M.,  White S. D.~M.,  Jenkins A.,   Pearce F.~R.,  1999, \mn@doi
  [Monthly Notices of the Royal Astronomical Society]
  {10.1046/j.1365-8711.1999.02400.x}, 308, 593

\bibitem[\protect\citeauthoryear{Cooray \& Sheth}{Cooray \&
  Sheth}{2002}]{Cooray:2002-Halo-models-of-large-scale-str}
Cooray A.,  Sheth R.,  2002, \mn@doi [Physics Reports-Review Section of Physics
  Letters] {10.1016/S0370-1573(02)00276-4}, 372, 1

\bibitem[\protect\citeauthoryear{Faber \& Jackson}{Faber \&
  Jackson}{1976}]{Faber:1976-Velocity-Dispersions-and-Mass-}
Faber S.~M.,  Jackson R.~E.,  1976, \mn@doi [Astrophysical Journal]
  {10.1086/154215}, 204, 668

\bibitem[\protect\citeauthoryear{Frenk et~al.,}{Frenk
  et~al.}{2000}]{Frenk:2000-Public-Release-of-N-body-simul}
Frenk C.~S.,  et~al., 2000, \mn@doi [arXiv:astro-ph/0007362v1]
  {10.48550/arXiv.astro-ph/0007362}

\bibitem[\protect\citeauthoryear{Girelli, Pozzetti, Bolzonella, Giocoli,
  Marulli  \& Baldi}{Girelli
  et~al.}{2020}]{Girelli:2020-The-stellar-to-halo-mass-relat}
Girelli G.,  Pozzetti L.,  Bolzonella M.,  Giocoli C.,  Marulli F.,   Baldi M.,
   2020, \mn@doi [Astronomy \& Astrophysics] {10.1051/0004-6361/201936329}, 634

\bibitem[\protect\citeauthoryear{Gunn \& Gott}{Gunn \&
  Gott}{1972}]{Gunn:1972-Infall-of-Matter-into-Clusters}
Gunn J.~E.,  Gott J.~R.,  1972, \mn@doi [Astrophysical Journal]
  {10.1086/151605}, 176, 1

\bibitem[\protect\citeauthoryear{Guo, White, Li  \& Boylan-Kolchin}{Guo
  et~al.}{2010}]{Guo:2010-How-do-galaxies-populate-dark-}
Guo Q.,  White S.,  Li C.,   Boylan-Kolchin M.,  2010, \mn@doi [Monthly Notices
  of the Royal Astronomical Society] {10.1111/j.1365-2966.2010.16341.x}, 404,
  1111

\bibitem[\protect\citeauthoryear{Jenkins et~al.,}{Jenkins
  et~al.}{1998}]{Jenkins:1998-Evolution-of-structure-in-cold}
Jenkins A.,  et~al., 1998, \mn@doi [Astrophysical Journal] {10.1086/305615},
  499, 20

\bibitem[\protect\citeauthoryear{{Lelli}, {McGaugh}  \& {Schombert}}{{Lelli}
  et~al.}{2016}]{Lelli:2016-SPARC-Mass-Models-for-175-Disk-Galaxies}
{Lelli} F.,  {McGaugh} S.~S.,   {Schombert} J.~M.,  2016, \mn@doi [\aj]
  {10.3847/0004-6256/152/6/157}, \href
  {https://ui.adsabs.harvard.edu/abs/2016AJ....152..157L} {152, 157}

\bibitem[\protect\citeauthoryear{Lelli, McGaugh, Schombert, Desmond  \&
  Katz}{Lelli et~al.}{2019}]{Lelli:2019-The-baryonic-Tully-Fisher-rela}
Lelli F.,  McGaugh S.~S.,  Schombert J.~M.,  Desmond H.,   Katz H.,  2019,
  \mn@doi [Monthly Notices of the Royal Astronomical Society]
  {10.1093/mnras/stz205}, 484, 3267

\bibitem[\protect\citeauthoryear{McGaugh \& de Blok}{McGaugh \&
  de~Blok}{1998}]{McGaugh:1998-Testing-the-dark-matter-hypoth}
McGaugh S.~S.,  de Blok W. J.~G.,  1998, \mn@doi [Astrophysical Journal]
  {10.1086/305612}, 499, 41

\bibitem[\protect\citeauthoryear{McGaugh, Schombert, Bothun  \& de
  Blok}{McGaugh et~al.}{2000}]{McGaugh:2000-The-baryonic-Tully-Fisher-rela}
McGaugh S.~S.,  Schombert J.~M.,  Bothun G.~D.,   de Blok W. J.~G.,  2000,
  \mn@doi [Astrophysical Journal] {10.1086/312628}, 533, L99

\bibitem[\protect\citeauthoryear{Milgrom}{Milgrom}{1983}]{Milgrom:1983-A-Modification-of-the-Newtonia}
Milgrom M.,  1983, \mn@doi [Astrophysical Journal] {10.1086/161130}, 270, 365

\bibitem[\protect\citeauthoryear{Mo, van~den Bosch  \& White}{Mo
  et~al.}{2010}]{Mo:2010-Galaxy-formation-and-evolution}
Mo H.,  van~den Bosch F.,   White S.,  2010, Galaxy formation and evolution.
Cambridge University Press, Cambridge

\bibitem[\protect\citeauthoryear{Moster, Somerville, Maulbetsch, van~den Bosch,
  Maccio, Naab  \& Oser}{Moster
  et~al.}{2010}]{Moster:2010-Constraints-on-the-Relationshi}
Moster B.~P.,  Somerville R.~S.,  Maulbetsch C.,  van~den Bosch F.~C.,  Maccio
  A.~V.,  Naab T.,   Oser L.,  2010, \mn@doi [Astrophysical Journal]
  {10.1088/0004-637x/710/2/903}, 710, 903

\bibitem[\protect\citeauthoryear{Moster, Naab  \& White}{Moster
  et~al.}{2013}]{Moster:2013-Galactic-star-formation-and-ac}
Moster B.~P.,  Naab T.,   White S. D.~M.,  2013, \mn@doi [Monthly Notices of
  the Royal Astronomical Society] {10.1093/mnras/sts261}, 428, 3121

\bibitem[\protect\citeauthoryear{Neyman \& Scott}{Neyman \&
  Scott}{1952}]{Neyman:1952-A-Theory-of-the-Spatial-Distri}
Neyman J.,  Scott E.~L.,  1952, \mn@doi [Astrophysical Journal]
  {10.1086/145599}, 116, 144

\bibitem[\protect\citeauthoryear{Peebles}{Peebles}{1980}]{Peebles:1980-The-Large-Scale-Structure-of-t}
Peebles P. J.~E.,  1980, The Large-Scale Structure of the Universe.
Princeton University Press, Princeton, NJ

\bibitem[\protect\citeauthoryear{Press \& Schechter}{Press \&
  Schechter}{1974}]{Press:1974-Formation-of-Galaxies-and-Clus}
Press W.~H.,  Schechter P.,  1974, \mn@doi [Astrophysical Journal]
  {10.1086/152650}, 187, 425

\bibitem[\protect\citeauthoryear{Read \& Trentham}{Read \&
  Trentham}{2005}]{Read:2005-The-baryonic-mass-function-of-}
Read J.~I.,  Trentham N.,  2005, \mn@doi [Philosophical Transactions of the
  Royal Society a-Mathematical Physical and Engineering Sciences]
  {10.1098/rsta.2005.1648}, 363, 2693

\bibitem[\protect\citeauthoryear{Richardson}{Richardson}{1922}]{Richardson:1922-Weather-Prediction-by-Numerica}
Richardson L.~F.,  1922, Weather Prediction by Numerical Process.
Cambridge University Press, Cambridge, UK

\bibitem[\protect\citeauthoryear{Rubin \& Ford}{Rubin \&
  Ford}{1970}]{Rubin:1970-Rotation-of-Andromeda-Nebula-f}
Rubin V.~C.,  Ford W.~K.,  1970, \mn@doi [Astrophysical Journal]
  {10.1086/150317}, 159, 379

\bibitem[\protect\citeauthoryear{Rubin, Ford  \& Thonnard}{Rubin
  et~al.}{1980}]{Rubin:1980-Rotational-Properties-of-21-Sc}
Rubin V.~C.,  Ford W.~K.,   Thonnard N.,  1980, \mn@doi [Astrophysical Journal]
  {10.1086/158003}, 238, 471

\bibitem[\protect\citeauthoryear{Sheth, Mo  \& Tormen}{Sheth
  et~al.}{2001}]{Sheth:2001-Ellipsoidal-collapse-and-an-im}
Sheth R.~K.,  Mo H.~J.,   Tormen G.,  2001, \mn@doi [Monthly Notices of the
  Royal Astronomical Society] {10.1046/j.1365-8711.2001.04006.x}, 323, 1

\bibitem[\protect\citeauthoryear{Taylor}{Taylor}{1935}]{Taylor:1935-Statistical-theory-of-turbulan}
Taylor G.~I.,  1935, \mn@doi [Proceedings of the royal society A]
  {10.1098/rspa.1935.0158}, 151, 421

\bibitem[\protect\citeauthoryear{Taylor}{Taylor}{1938}]{Taylor:1938-Production-and-dissipation-of-}
Taylor G.~I.,  1938, \mn@doi [Proceedings of the Royal Society of London Series
  a-Mathematical and Physical Sciences] {10.1098/rspa.1938.0002}, 164, 0015

\bibitem[\protect\citeauthoryear{Tully \& Fisher}{Tully \&
  Fisher}{1977}]{Tully:1977-New-Method-of-Determining-Dist}
Tully R.~B.,  Fisher J.~R.,  1977, Astronomy \& Astrophysics, 54, 661

\bibitem[\protect\citeauthoryear{Wechsler \& Tinker}{Wechsler \&
  Tinker}{2018}]{Wechsler:2018-The-Connection-Between-Galaxie}
Wechsler R.~H.,  Tinker J.~L.,  2018, \mn@doi [Annual Review of Astronomy and
  Astrophysics, Vol 56] {10.1146/annurev-astro-081817-051756}, 56, 435

\bibitem[\protect\citeauthoryear{Xu}{Xu}{2021a}]{Xu:2021-Inverse-mass-cascade-mass-function}
Xu Z.,  2021a, \mn@doi [arXiv e-prints] {10.48550/ARXIV.2109.09985}, p.
  arXiv:2109.09985

\bibitem[\protect\citeauthoryear{Xu}{Xu}{2021b}]{Xu:2021-Inverse-mass-cascade-halo-density}
Xu Z.,  2021b, \mn@doi [arXiv e-prints] {10.48550/ARXIV.2109.12244}, p.
  arXiv:2109.12244

\bibitem[\protect\citeauthoryear{Xu}{Xu}{2021c}]{Xu:2021-The-maximum-entropy-distributi}
Xu Z.,  2021c, \mn@doi [arXiv e-prints] {10.48550/ARXIV.2110.03126}, p.
  arXiv:2110.03126

\bibitem[\protect\citeauthoryear{Xu}{Xu}{2021d}]{Xu:2021-A-non-radial-two-body-collapse}
Xu Z.,  2021d, \mn@doi [arXiv e-prints] {10.48550/ARXIV.2110.05784}, p.
  arXiv:2110.05784

\bibitem[\protect\citeauthoryear{Xu}{Xu}{2021e}]{Xu:2021-Mass-functions-of-dark-matter-}
Xu Z.,  2021e, \mn@doi [arXiv e-prints] {10.48550/ARXIV.2110.09676}, p.
  arXiv:2110.09676

\bibitem[\protect\citeauthoryear{Xu}{Xu}{2021f}]{Xu:2021-Inverse-and-direct-cascade-of-}
Xu Z.,  2021f, \mn@doi [arXiv e-prints] {10.48550/ARXIV.2110.13885}, p.
  arXiv:2110.13885

\bibitem[\protect\citeauthoryear{Xu}{Xu}{2022a}]{Xu:2022-Dark_matter-flow-and-hydrodynamic-turbulence-presentation}
Xu Z.,  2022a, A comparative study of dark matter flow \& hydrodynamic
  turbulence and its applications, \mn@doi{10.5281/zenodo.6569901}, \url
  {http://dx.doi.org/10.5281/zenodo.6569901}

\bibitem[\protect\citeauthoryear{Xu}{Xu}{2022d}]{Xu:Dark_matter_flow_dataset_2022_all_files}
Xu Z.,  2022d, Dark matter flow dataset, \mn@doi{10.5281/zenodo.6586212}, \url
  {https://github.com/ZhijieXu2022/dark\_matter\_flow\_dataset}

\bibitem[\protect\citeauthoryear{Xu}{Xu}{2022b}]{Xu:2022-Dark_matter-flow-dataset-part1}
Xu Z.,  2022b, Dark matter flow dataset Part I: Halo-based statistics from
  cosmological N-body simulation, \mn@doi{10.5281/zenodo.6541230}, \url
  {http://dx.doi.org/10.5281/zenodo.6541230}

\bibitem[\protect\citeauthoryear{Xu}{Xu}{2022c}]{Xu:2022-Dark_matter-flow-dataset-part2}
Xu Z.,  2022c, Dark matter flow dataset Part II: Correlation-based statistics
  from cosmological N-body simulation, \mn@doi{10.5281/zenodo.6569898}, \url
  {http://dx.doi.org/10.5281/zenodo.6569898}

\bibitem[\protect\citeauthoryear{Xu}{Xu}{2022e}]{Xu:2022-The-mean-flow--velocity-disper}
Xu Z.,  2022e, \mn@doi [arXiv e-prints] {10.48550/ARXIV.2201.12665}, p.
  arXiv:2201.12665

\bibitem[\protect\citeauthoryear{Xu}{Xu}{2022f}]{Xu:2022-The-statistical-theory-of-2nd}
Xu Z.,  2022f, \mn@doi [arXiv e-prints] {10.48550/ARXIV.2202.00910}, p.
  arXiv:2202.00910

\bibitem[\protect\citeauthoryear{Xu}{Xu}{2022g}]{Xu:2022-The-statistical-theory-of-3rd}
Xu Z.,  2022g, \mn@doi [arXiv e-prints] {10.48550/ARXIV.2202.02991}, p.
  arXiv:2202.02991

\bibitem[\protect\citeauthoryear{Xu}{Xu}{2022h}]{Xu:2022-The-evolution-of-energy--momen}
Xu Z.,  2022h, \mn@doi [arXiv e-prints] {10.48550/ARXIV.2202.04054}, p.
  arXiv:2202.04054

\bibitem[\protect\citeauthoryear{Xu}{Xu}{2022i}]{Xu:2022-Two-thirds-law-for-pairwise-ve}
Xu Z.,  2022i, \mn@doi [arXiv e-prints] {10.48550/ARXIV.2202.06515}, p.
  arXiv:2202.06515

\bibitem[\protect\citeauthoryear{Xu}{Xu}{2022j}]{Xu:2022-Postulating-dark-matter-partic}
Xu Z.,  2022j, \mn@doi [arXiv e-prints] {10.48550/ARXIV.2202.07240}, p.
  arXiv:2202.07240

\bibitem[\protect\citeauthoryear{Xu}{Xu}{2022k}]{Xu:2022-The-origin-of-MOND-acceleratio}
Xu Z.,  2022k, \mn@doi [arXiv e-prints] {10.48550/ARXIV.2203.05606}, p.
  arXiv:2203.05606

\bibitem[\protect\citeauthoryear{de Karman \& Howarth}{de~Karman \&
  Howarth}{1938}]{de_Karman:1938-On-the-statistical-theory-of-i}
de Karman T.,  Howarth L.,  1938, \mn@doi [Proceedings of the Royal Society of
  London Series a-Mathematical and Physical Sciences] {10.1098/rspa.1938.0013},
  164, 0192

\makeatother
\end{thebibliography}

\appendix

\label{lastpage}
\end{document}